\documentclass[reprint, amsmath, amssymb, showpacs, floatfix, longbibliography, aps, twocolumn, superscriptaddress, footinbib]{revtex4-1}
\usepackage{CJKutf8}

\usepackage{amsthm}

\usepackage[utf8]{inputenc}
\usepackage{amsmath,amssymb,amsfonts,stmaryrd}
\usepackage{physics}
\usepackage{dsfont}
\usepackage{graphicx}
\usepackage[dvipsnames]{xcolor}
\usepackage{graphicx}
\usepackage{cancel}
\usepackage{natbib}
\usepackage{color}
\usepackage{tikz}
\usepackage{mathrsfs}
\usepackage{relsize}
\usepackage{multirow}
\usepackage[linktocpage]{hyperref}
\usepackage[all]{xy}
\usepackage[export]{adjustbox}
\hypersetup{colorlinks=true,citecolor=blue,linkcolor=blue, urlcolor=blue, breaklinks=true}

\usepackage{float}

\usepackage{bm} 
\usepackage{subfigure} 
\usepackage{environ}
\usepackage{url}
\usepackage[margin=0.75in]{geometry}
\usepackage{ulem}

\usepackage{mathtools}

\newcommand{\D}{{\mathbb{D}}}
\newcommand{\lX}{{\overrightarrow{X}}}
\newcommand{\rX}{{\overleftarrow{X}}}

\newcommand{\Z}{{\mathbb Z}}

\newcommand{\eps}{\epsilon}

\NewEnviron{eqs}{%
\begin{equation}\begin{split}
    \BODY
\end{split}\end{equation}}

\definecolor{dkgreen}{rgb}{0,0.5,0}

\theoremstyle{definition}

\theoremstyle{remark}

\def\mM{\mathcal{M}}
\def\mU{\mathcal{U}}

\begin{document}

\begin{CJK*}{UTF8}{bsmi}

\title{Invariants of Sequential Circuits and Generalized Non-Abelian Statistics}

\author{Shintaro Sato}
\affiliation{Department of Physics, The University of Tokyo, 7-3-1 Hongo, Bunkyo-ku, Tokyo 113-0033, Japan}

\author{Yoshimasa Hidaka}
\affiliation{Yukawa Institute for Theoretical Physics, Kyoto University, Kyoto 606-8502, Japan}
\affiliation{RIKEN Center for Interdisciplinary Theoretical and Mathematical Sciences (iTHEMS), RIKEN, Wako 351-0198, Japan}

\author{Ryohei Kobayashi}
\email[E-mail: ]{ryohei.k@ap.t.u-tokyo.ac.jp}
\affiliation{Department of Applied Physics, The University of Tokyo, Tokyo 113-8656, Japan}

\date{\today}
\preprint{RIKEN-iTHEMS-Report-26}
\begin{abstract}
Non-invertible symmetries in quantum many-body systems generally give rise to sequential unitary circuits that move symmetry defects. In this paper, we investigate invariants defined by sequences of such circuits, which move non-invertible defects and generate a Berry phase evaluated on quantum states with defects. We show that this Berry phase generally defines an invariant under local deformations, provided that the sequential circuits preserve the locality of those deformations. This invariant also rules out a short-range-entangled state that preserves the non-invertible symmetry, thereby signaling the 't Hooft anomaly of a non-invertible symmetry purely in terms of unitary operators acting on a state.
We then apply this framework to loop excitations in three spatial dimensions and identify a new loop excitation in the (3+1)D $\mathbb{D}_4$ topological order, which we dub a non-Abelian fermionic loop. Using the invariant of sequential circuits, we characterize the statistics of non-Abelian fermionic loops. In addition, we find a new (3+1)D mixed topological order with a single non-Abelian fermionic loop, whose long-range entanglement is protected by an invariant of sequential circuits.
 \end{abstract}

\maketitle

\end{CJK*}




\section{Introduction}
Symmetry plays a central role in quantum many-body physics, from organizing phases of matter to constraining low-energy dynamics. In particular, global symmetries can protect nontrivial phases and impose powerful selection rules on excitations and responses \cite{Chen:2011pg, Gaiotto:2014kfa}. More recently, non-invertible symmetries have emerged as a broader class of generalized symmetries that extend the conventional group symmetries and arise naturally in quantum field theory and quantum matter~\cite{Chang:2018iay,Aasen:2020jwb,Choi:2021kmx,Kaidi:2021xfk,Choi:2022zal,Zhang:2023wlu,Cordova:2023bja,Bhardwaj:2023ayw,Bhardwaj:2023fca, schafernameki2023ictp,Shao:2023gho}.

In quantum many-body systems on the lattice, however, a general definition of non-invertible symmetry still remains under debate.
Unlike invertible global symmetries, which are represented by locality-preserving unitary operators, non-invertible symmetries are not generically captured by a single unitary acting on the whole Hilbert space.
The action of non-invertible symmetry in QFT is rather described by a quantum channel \cite{okada2024noninvertible, bartsch2026wignernoninvertiblesymmetriespreserve}, hence lattice non-invertible symmetries should be defined through locality-preserving quantum channels with certain additional conditions \cite{jones2026structurecategoricaldualityoperators, noninvertibleQCA}. See Ref.~\cite{evans2026operatoralgebraicapproachfusion, inamura2026remarksnoninvertiblesymmetriestensor, wen2026noninvertiblesymmetriestensorproducthilbert} for recent discussions on locality-preserving non-invertible symmetry operators on lattice models.
In this work we do not attempt to directly answer this foundational question. Instead, we focus on a robust feature that non-invertible symmetry defects are moved by sequential unitary circuits \cite{tantivasadakarn2025sequential}. This property generally holds in lattice realizations and provides a simple, tractable handle on symmetries without directly using non-invertible symmetry operators or quantum channels.

A key subtlety is that sequential circuits are not locality-preserving automorphisms of the full local operator algebra. Accordingly, an internal non-invertible symmetry action is typically realized by combining a sequential circuit with a projection onto a subspace (or subalgebra) whose locality structure is preserved under the symmetry action; the projection is what renders the action non-invertible. Familiar examples such as the Kramers-Wannier duality fit naturally into this pattern~\cite{seifnashri2024cluster}. This viewpoint motivates us to treat sequential unitary circuits as the primitive objects for characterizing fundamental properties of symmetries.

One of the most important consequences of symmetry is the possibility of 't Hooft anomalies, which obstruct the realization of a symmetric short-range-entangled (SRE) state and strongly constrain low-energy dynamics both in continuum QFT~\cite{shimizu2018,hsin2020se,seifnashri2021sym} and lattice systems~\cite{chen2013SPT,ElseNayak2014,tiwari2018, LSM1961, oshikawa2, hastings2004,cheng2016set, cho2017anomaly, Kobayashi2019lsm, else2020lsm, prem2020lsm, seiberg2022lsm, Seifnashri2024lsm}. For invertible symmetries, there has been substantial progress in formulating and diagnosing anomalies directly in lattice systems \cite{ElseNayak2014, Feng2026higher, Kobayashi2026generalized,
kawagoe2025anomaly, kapustin2025anomaly2d, shirley2025QCA, tu2025anomaliesglobalsymmetrieslattice}. By contrast, lattice diagnostics of anomalies for non-invertible symmetries remain comparatively underdeveloped, in part because the definition of non-invertible symmetry on lattices itself is still not fully settled. This motivates the indicator of anomalies that can be defined directly from concrete sequential unitaries that move defects.

In this paper, we introduce such an indicator that characterizes a subclass of non-invertible symmetry anomalies: an invariant associated with a sequence of sequential circuits moving non-invertible defects. Acting on a family of quantum states with defects, this sequence generates a Berry phase. We show that this Berry phase is invariant under local deformations, provided that the non-invertible symmetry action preserves the locality of those deformations.
We generally show that this invariant forbids a symmetric SRE realization. As a result, it directly signals an 't Hooft anomaly of non-invertible symmetries purely in terms of unitary operators acting on states, without referring to an explicit non-invertible symmetry operator on the full Hilbert space. This formulation is regarded as non-invertible analogue of the ``generalized statistics'' developed in Ref.~\cite{Kobayashi2026generalized}, at the same time higher-dimensional generalizations of braiding invariants in (2+1)D \cite{kawagoe2020microscopic}, which are the Berry phase invariants of quantum states with defect insertions.
This provides a practical and physically transparent characterization of symmetry response and dynamical constraints of non-invertible symmetries.

As an application of our framework, we identify a new loop excitation in (3+1)D $\mathbb{D}_4$ topological order (where $\mathbb{D}_4$ is the dihedral group of order 8), which we call a non-Abelian fermionic loop. 
This is a non-Abelian analogue of the fermionic loop excitations found in $\Z_2$ gauge theories \cite{Thorngren2015framed,  Fidkowski:2021unr, Chen:2021xks, Kobayashi2026generalized, feng2026paulistabilizerformalismtopological}.
We characterize its statistics through the sequential-circuit invariant, and show that this non-Abelian loop excitation exhibits $\mathbb{Z}_2$ fermionic self-statistics. 

Recently, rapid progress in the study of mixed-state phases~\cite{Sohal:2024qvq,Ellison:2024svg,Wang2025intrinsic,zhang2024, Lee2023criticality, Fan2024diagnostics, bao2023mixed, chen2024sep,sun2025anomalousMPO} has clarified how the anomalies of strong and weak 1-form symmetries lead to nontrivial entanglement patterns in mixed states beyond those in ground states of gapped local Hamiltonians. In particular, the anomalies of strong higher-form symmetries generally lead to long-range entanglement that protects nontrivial mixed state phases~\cite{Lessa:2024wcw,Hsin:2023jqa,Zhou:2023icb,Zang:2023qou,Wang:2024vjl,Sohal:2024qvq,Ellison:2024svg,chirame2024to,xu2025avg,zhang2024,zhou2025finiteT,lessa2025higher, li2025much, hsin2025higherformanomaliesimplyintrinsic}, referred to as intrinsic long-range entanglement.
As an application of non-Abelian statistics described in this paper, we identify a new (3+1)D mixed topological order containing a single non-Abelian fermionic loop, whose long-range entanglement is protected by this invariant.

This paper is organized as follows. In Sec.~\ref{sec:invariance}, we describe a general framework for the Berry phase invariants of sequential unitary circuits.
In Sec.~\ref{sec:anomaly}, we demonstrate that the invariant generally obstructs the realization in SRE states. In Sec.~\ref{sec:floop}, we describe continuum and lattice descriptions of non-Abelian fermionic loop excitations in $\mathbb{D}_4$ topological order in (3+1)D. In Sec.~\ref{sec:mixed}, we describe the new intrinsically mixed topological order with a single non-Abelian fermionic loop excitation in (3+1)D. We conclude this paper with possible future directions in Sec.~\ref{sec:discussions}.

\section{Invariant of sequential circuits}
\label{sec:invariance}
We formulate an invariant of sequential circuits associated with non-invertible symmetry, under a set of physically motivated assumptions that the sequential unitaries are required to satisfy.
We consider a tensor-product Hilbert space in a generic spatial dimension $d$. In the presence of a non-invertible symmetry, one can generally define a defect state $\ket{\mathcal{D}}$ with an inserted defect network $\mathcal D$. Our goal is to construct a Berry phase from a family of defect states.

\begin{figure}[htb]
    \centering
    \includegraphics[width=0.5\linewidth]{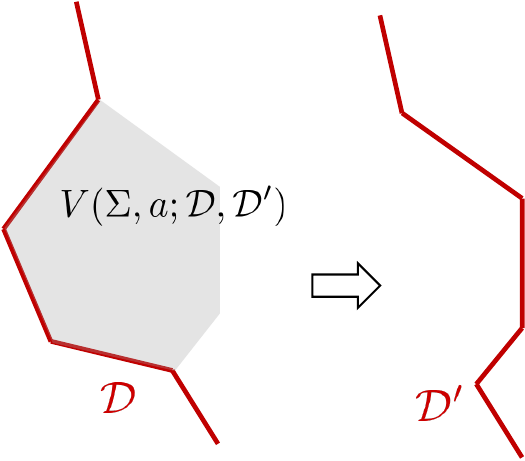}
    \caption{A sequential circuit $V(\Sigma,a;\mathcal{D},\mathcal{D}')$ is supported at a region $\Sigma$. The corresponding symmetry generator is labeled by $a$ (generally an object of a symmetry category), and transforms a defect network. }
    \label{fig:moving}
\end{figure}

We begin by fixing a finite set of defect-network configurations $\{\mathcal{D}\}$ in $d$-dimensional space. Each defect network $\mathcal D$ is associated with a quantum state $\ket{\mathcal{D}}$ carrying the corresponding defect insertion. For each pair $(\mathcal{D},\mathcal{D}')$, we further fix a defect-``movement'' operator $V(\Sigma,a;\mathcal{D},\mathcal{D}')$, defined by
$V(\Sigma,a;\mathcal D,\mathcal D')\ket{\mathcal D}
= e^{i\theta(\Sigma,a;\mathcal D,\mathcal D')}\ket{\mathcal D'}$ (see Fig.~\ref{fig:moving}).
Here, $\Sigma$ denotes the support of the operator, and $a$ labels a simple object in the symmetry (generally a higher fusion) category, representing a generator of the non-invertible symmetry. We assume that each movement operator $V(\Sigma,a;\mathcal D,\mathcal D')$ is implemented by a sequential unitary circuit. The maximal geometrical range of each local circuit in the sequential unitary is denoted by $\epsilon$.

We then consider a Berry phase associated with a sequence of defect-movement operators acting on a fixed defect state:
\begin{align}
   & \bra{\mathcal{D}_1}\left(\prod_{j=1}^{n-1}V(\Sigma_j,a_j;\mathcal{D}_j,\mathcal{D}_{j+1}) \right)\ket{\mathcal{D}_1} \nonumber \\
   &= \exp\left( i\sum_{j=1}^{n-1} \theta(\Sigma_j,a_j;\mathcal{D}_j,\mathcal{D}_{j+1}) \right)
   \label{eq:berryphase}
 \end{align}
Not every such sequence defines an invariant under local deformations of the operators and states. To formulate and establish invariance under admissible deformations, we impose several conditions on the movement operators and on the allowed sequences of their composition:

\begin{itemize}
\item \textit{Fixed operator support. }
For a given pair of defect configurations $\mathcal{D}, \mathcal{D}'$ together with a label of the movement operator $a$, the support $\Sigma$ of the operator $V(\Sigma,a;\mathcal{D},\mathcal{D}')$ is uniquely fixed. Namely, if two operators $V(\Sigma,a;\mathcal{D},\mathcal{D}'),V(\Sigma',a;\mathcal{D},\mathcal{D}')$ appear in the sequence \eqref{eq:berryphase}, then $\Sigma=\Sigma'$. Therefore, we sometimes write $V(\Sigma,a;\mathcal{D},\mathcal{D}')$ as $V(a;\mathcal{D},\mathcal{D}')$ by omitting $\Sigma$.

\item \textit{Local uniqueness of circuits. }
Consider a point $p$ of $\Sigma$. Then, we require that the sequential circuit $V(\Sigma,a;\mathcal{D},\mathcal{D}')$ can be expressed in the form of
\begin{align}
    V(\Sigma,a;\mathcal{D},\mathcal{D}') = U' U_{r,p} U''
    \label{eq:decompose V}
\end{align}
where $U_{r,p}$ is a circuit supported on a ball $D_{r,p}$ of $p$ with radius $r$, and the supports of $U'$ and $U''$ are away from $p$ (at a distance $r-\eps$). Here, $r$ is taken much larger than $\epsilon$, while much smaller than the size of $V(\Sigma,a;\mathcal{D},\mathcal{D}')$.
Further, consider movement operators $V(\Sigma,a;\mathcal{D}_1,\mathcal{D}_2)$, $V(\Sigma,a;\mathcal{D}_3,\mathcal{D}_4)$, where the restrictions of the defect configurations are identical in the $r$-neighborhood of $p$: $(\mathcal{D}_1,\mathcal{D}_2)|_{D_{r,p}} = (\mathcal{D}_3,\mathcal{D}_4)|_{D_{r,p}}$. Then, we require that the circuit $U_{r,p}$ in the $r$-neighborhood of $p$ can be taken to be identical: 
\begin{align}
    U_{r,p}(a;\mathcal{D}_1,\mathcal{D}_2)=U_{r,p}(a;\mathcal{D}_3,\mathcal{D}_4)~.
    \label{eq:locality of V}
\end{align}
In other words, the local circuit $U_{r,p}$ depends on defect configurations only through those in the vicinity of $p$: $(\mathcal{D}_1,\mathcal{D}_2)|_{D_{r,p}}$.

\item \textit{Inverse of sequential circuits.}
For any pair of defect configurations $\mathcal{D}, \mathcal{D}'$ and an operator label $a$, we require
    \begin{align}
 V(\Sigma,a^{*}; \mathcal{D}',\mathcal{D})=V(\Sigma,a; \mathcal{D},\mathcal{D}')^\dagger~,
 \label{eq: V inverse}
 \end{align}
 where $a^*$ is a dual of $a$ \footnote{Here we assumed that the left and right dual of the symmetry fusion $d$-category are identical. That generally requires additional structure to the symmetry category, e.g., being pivotal. }, hence $\theta(\Sigma,a^*;\mathcal{D},\mathcal{D}')+\theta(\Sigma,a;\mathcal{D}',\mathcal{D})=0$ mod $2\pi$. 

 \item \textit{Defect creation operator.} 
 Each state is obtained by acting a defect creation operator on the canonical state $\ket{0}$ with an empty defect configuration:
 \begin{align}
     \ket{\mathcal{D}} = V(\mathcal{D})\ket{0}~.
 \end{align}
The operator $V(\mathcal{D})$ is not required to be unitary.
 
\end{itemize}

The Berry phase is expressed as a sum of phase factors
\begin{align}
    \Theta = \sum_{a,\mathcal{D},\mathcal{D}'} \eps(\Sigma,a;\mathcal{D}, \mathcal{D}') \theta(\Sigma,a;\mathcal{D}, \mathcal{D}')~,
\end{align}
with $\eps(\Sigma,a;\mathcal{D}, \mathcal{D}')\in\Z$. Then, it turns out that the necessary and sufficient condition for $\Theta$ to define an invariant is expressed as a set of linear constraints satisfied by the integral coefficients $\epsilon$:
\begin{enumerate}
    \item By redefining each state by a phase $\ket{\mathcal{D}}\to e^{i\phi(\mathcal{D})}\ket{\mathcal{D}}$, each phase factor is shifted as $\theta(\Sigma,a;\mathcal{D},\mathcal{D}')\to\theta(\Sigma,a;\mathcal{D},\mathcal{D}')+\phi(\mathcal{D})-\phi(\mathcal{D}')$. Invariance under these shifts corresponds to the linear constraints
    \begin{align}
        \sum_{a,\mathcal{D}'} \eps(\Sigma,a;\mathcal{D},\mathcal{D}') - \sum_{a,\mathcal{D}'} \eps(\Sigma,a^*;\mathcal{D}',\mathcal{D}) = 0~.
        \label{eq:redefining constraint}
        \end{align}
        
    \item We can redefine each operator by a phase $V(\Sigma,a;\mathcal{D},\mathcal{D}')\to e^{i\phi(\Sigma,a;\mathcal{D},\mathcal{D}')}V(\Sigma,a;\mathcal{D},\mathcal{D}')$. The redefinitions have to be consistent with the constraints \eqref{eq:locality of V}, \eqref{eq: V inverse}; for instance, when $\Sigma_{\mathcal{D}_1, \mathcal{D}_2} =\Sigma_{\mathcal{D}_3, \mathcal{D}_4}=\Sigma$ and $(\mathcal{D}_1,\mathcal{D}_2)|_{\partial\Sigma} =(\mathcal{D}_3,\mathcal{D}_4)|_{\partial\Sigma}$, the phase shift must be identical: $\phi(\Sigma,a;\mathcal{D}_1,\mathcal{D}_2) = \phi(\Sigma,a;\mathcal{D}_3,\mathcal{D}_4)$. The invariance under such phase redefinitions corresponds to the linear constraints
\begin{align}
        \sum_{\substack{\mathcal{D}_1,\mathcal{D}_2 \\ ( \mathcal{D}_1,\mathcal{D}_2)|_{\partial\Sigma}= (\mathcal{D}_1|, \mathcal{D}_2|)}} \left(\eps(a;\mathcal{D}_1, \mathcal{D}_2) -  \eps(a^*;\mathcal{D}_2,\mathcal{D}_1 )\right) = 0~,
        \end{align}
        where we omitted $\Sigma$ from the expression since it is uniquely determined by $(a,\mathcal{D}_1,\mathcal{D}_2)$, and the sum means that we sum over defects with fixed configurations $(\mathcal{D}_{1}|, \mathcal{D}_{2}|)$ at $\partial\Sigma$. 
        \item 
    We also require the Berry phase to be invariant under admissible local deformations of a movement operator $V(\Sigma,a;\mathcal{D}_1,\mathcal{D}_2)$. Such deformations may occur near the boundary $\partial\Sigma$ and are associated with possible redefinitions of defect states along the defect network. Unlike the case of invertible symmetries, however, local deformations of sequential circuits involve an additional subtlety.
To see this, consider a sequential-circuit representation of $V(\Sigma,a;\mathcal{D}_1,\mathcal{D}_2)$ of the form $V(\Sigma,a;\mathcal{D}_1,\mathcal{D}_2)=\cdots U' U_{\epsilon,p} U'' \cdots$, where $U_{\eps,p}$ is a local circuit supported at $p$ near the boundary $\partial \Sigma$. A local deformation at a point $p\in\partial\Sigma$ is implemented by modifying the local circuit $U_{\epsilon,p}$ as $U_{\epsilon,p}\to U'_{\epsilon,p}=V_{\epsilon,p}U_{\epsilon,p}$, where $V_{\epsilon,p}$ is a local circuit supported in a region of size $\epsilon$. However, this seemingly local modification can induce a nonlocal change in the Berry phase, because a sequential circuit does not generally preserve operator locality; the conjugation action of $U'$ or $U''$ on $U_{\eps,p}$ can become nonlocal. The induced redefinition of $V(\Sigma,a;\mathcal{D}_1,\mathcal{D}_2)$ may hence become nonlocal, so invariance of the Berry phase is not expected in general.

For this reason, we restrict to local deformations for which the locality of $V_{\epsilon,p}$ is preserved under the action of the symmetry (movement) operators.
As an example, when the non-invertible symmetry is the $\mathbb{Z}_2$ Kramers-Wannier duality, we consider deformations that preserve the $\mathbb{Z}_2$ symmetry, since the duality preserves the locality of $\mathbb{Z}_2$-even operators.

In Appendix \ref{sec:berry_phases}, we show that invariance under such admissible local deformations is equivalent to a set of linear constraints on the integer coefficients ${\epsilon}$ that define the Berry phase,
\begin{align}
    \sum_{\substack{\mathcal{D}_1,\mathcal{D}_2, \\ (\mathcal{D}_1,\mathcal{D}_2)|_{D_{r,p}} = (\mathcal{D}_{1}|, \mathcal{D}_{2}|)}}(\epsilon(a;\mathcal{D}_1, \mathcal{D}_2)-\epsilon(a^*;\mathcal{D}_2, \mathcal{D}_1)) = 0~,
    \label{eq:epsilon locality constraint}
\end{align}
where we omitted $\Sigma$ from the expression since it is uniquely determined by $(a,\mathcal{D}_1,\mathcal{D}_2)$, and the sum means that we sum over defects with fixed configurations $(\mathcal{D}_{1}|, \mathcal{D}_{2}|)$ at the $r$-neighborhood of $p$. 
\end{enumerate}

\section{Berry phase as an 't Hooft anomaly} 
\label{sec:anomaly}

Here, we show that the generalized statistics invariant forbids a short-range-entangled (SRE) state preserving the non-invertible symmetry.
We demonstrate it by showing that the generalized statistics must have a trivial phase on an SRE state. This argument generalizes the discussions in Refs.~\cite{li2025much, Kobayashi2026generalized} to non-invertible defects with sequential circuits.

Assume that the symmetric ground state $|\Psi\rangle$ is SRE. 
Then there exists a finite-depth unitary circuit $W$ and a product state $|0\rangle$ such that
\begin{align}
|\Psi\rangle = W|0\rangle~, \label{eq:SRE_W_def}
\end{align}
with the defect state $\ket{\mathcal{D}}$ given by $\ket{\mathcal{D}}= V(\mathcal{D})\ket{\Psi}$.

Conjugating defect states and movement operators by $W$, we define
\begin{align}
\widetilde V(\Sigma,a;\mathcal D,\mathcal D') &:= W^\dagger V(\Sigma,a;\mathcal D,\mathcal D') W~, \cr
\widetilde V(\mathcal D)&:=W^\dagger V(\mathcal D)W~,\cr |\widetilde{\mathcal D}\rangle &:= W^\dagger|\mathcal D\rangle =\widetilde V(\mathcal D)\ket{0}~.
\label{eq:conjugated_definitions}
\end{align}
Note that $\widetilde{V}(\Sigma,a;\mathcal D,\mathcal D')$ is again a sequential unitary circuit.
Then any Berry phase of a closed sequence of movement operators is unchanged:
\begin{align}
&\bra{\mathcal D_1}\Bigl(\prod_{j=1}^{n-1} V(\Sigma_j,a_j;\mathcal D_j,\mathcal D_{j+1})\Bigr)\ket{\mathcal D_1}\cr
=&
\bra{\widetilde{\mathcal D}_1}\Bigl(\prod_{j=1}^{n-1} \widetilde V(\Sigma_j,a_j;\mathcal D_j,\mathcal D_{j+1})\Bigr)\ket{\widetilde{\mathcal D}_1}~.
\label{eq:berry_invariant_under_W}
\end{align}
Therefore, to analyze generalized-statistics invariants on SRE states, it suffices to work with the product reference state $|0\rangle$ and the conjugated operators $\widetilde V$.

\subsection{Factorization of defects in SRE states}

We show that, making a suitable topological-deformability assumption for defect creation operators $V(\mathcal{D})$, the above defect state $|\widetilde{\mathcal D}\rangle$ created on a product state forms a trivial product state away from the defect network. Concretely, we show that $|\widetilde{\mathcal D}\rangle$ has the form of 
\begin{align}
    |\widetilde{\mathcal D}\rangle = \ket{\psi}_A\otimes \ket{0}_{A^c}~,
    \label{eq:defect state is trivial}
\end{align}
where $A$ is an $\eps$-neighborhood of the defects $\mathcal{D}$, and $A^c$ is its complement, meaning that the defect state forms a product state away from the defects.

To see this, we first consider the case where in $d$ spatial dimensions, the movement operators $V(\Sigma,a;\mathcal{D},\mathcal{D}'),V(\Sigma',a;\mathcal{D},\mathcal{D}')$ have dimension $<d$, i.e., non-invertible higher-form symmetry. The case with the 0-form symmetry is discussed later.

We assume that the symmetry action is topological on defect states $\ket{\mathcal{D}}$: each movement operator $V(\Sigma,a;\mathcal{D},\mathcal{D}')$ can be topologically deformed so that its support $\Sigma$ is deformed into $\Sigma'$ with the same boundary $\partial\Sigma=\partial\Sigma'$, and the actions of $V(\Sigma,a;\mathcal{D},\mathcal{D}'),V(\Sigma',a;\mathcal{D},\mathcal{D}')$ on the defect states are identical as long as the deformation does not cross defects; $V(\Sigma,a;\mathcal{D},\mathcal{D}')\ket{\mathcal{D}} =V(\Sigma',a;\mathcal{D},\mathcal{D}')\ket{\mathcal{D}}$. We impose such a topological condition on the defect creation operators $V(\Sigma;\mathcal{D})$ as well.

Let $\Gamma(\mathcal D)$ denote the geometric support of the defect network.
Let $A$ be an $\eps$-neighborhood (a ``thickening'') of $\mathcal D$.
We write the Hilbert space as $\mathcal H=\mathcal H_A\otimes \mathcal H_{A^c}$.

Recall that defect states are obtained by acting with a defect-creation operator:
\begin{align}
|\widetilde{\mathcal D}\rangle = \widetilde V(\Sigma;\mathcal D)\,|0\rangle~, \label{eq:defect_state_from_creator}
\end{align}
where we write the support of the operator $\widetilde V(\mathcal{D})$ as $\Sigma$, such that the region $A$ is supported along the boundary of $\Sigma$.
Assuming that the symmetry operator is topological on the SRE state, there is another choice of symmetry operator $\widetilde V(\Sigma';\mathcal{D})$ supported at $\Sigma'$, such that
\begin{align}
    |\widetilde{\mathcal D}\rangle = \widetilde V(\Sigma';\mathcal D)\,|0\rangle~, \quad \Sigma\cap \Sigma'\subset A~.
    \label{eq:deformed defect creator}
\end{align}

Now, take a complement region of $A$ in the space and denote it by $A^c$. We write the projector onto the trivial state $\ket{0}$ within $A^c$ as $\Pi^{A^c}$. This projector is written in terms of projectors in the small regions,
\begin{align}
    \Pi^{A^c} = \prod_{L_j\in A^c} \Pi^{L_j}~,
\end{align}
where each region $L_j\subset A^c$ is small enough so that $\Pi^{L_j}$ commutes with either $\widetilde V(\Sigma;\mathcal{D})$ or $\widetilde V(\Sigma';\mathcal{D})$. Then, by using the expression $|\widetilde{\mathcal D}\rangle = \widetilde V(\Sigma;\mathcal D)\,|0\rangle =\widetilde V(\Sigma';\mathcal D)\,|0\rangle$, we get
\begin{align}
    \Pi^{A^c} |\widetilde{\mathcal D}\rangle = |\widetilde{\mathcal D}\rangle~,
\end{align}
implying that the defect state $|\widetilde{\mathcal D}\rangle$ has the form of $\ket{\psi}_A\otimes \ket{0}_{A^c}$, showing that the state factorizes into the state along the defect network and the trivial product state at the complement.

For the case of 0-form symmetry, when the space is a sphere $S^d$ and $A$ is a neighborhood of a $(d-1)$-sphere (an equator), one can assume the existence of $V(\Sigma';\mathcal{D})$ such that $\Sigma'$ is a complement of $\Sigma$ in the space, and $|{\mathcal D}\rangle =  V(\Sigma;\mathcal D)\,|0\rangle = V(\Sigma';\mathcal D)\,|0\rangle$ is satisfied. Then the above argument from Eq.~\eqref{eq:deformed defect creator} is valid for 0-form symmetry as well.

\subsection{Generalized statistics is trivial on factorized defect states}

We now show that, assuming the form of the defect state \eqref{eq:defect state is trivial}, the invariant of sequential circuits becomes trivial.

\subsubsection{When defects are points}

We first consider the case when the defects are 0d point objects; the defect state $\ket{\widetilde{\mathcal{D}}}$ has a local 0d state along the defect network,
\begin{align}
\ket{\widetilde{\mathcal{D}}} = \ket{\psi}_{\mathcal{D}}\otimes \ket{0}~,
\end{align}
where $\ket{\psi}_{\mathcal{D}}$ is regarded as a union of 0d local states localized at each defect.

For each 1d line operator support $\Sigma$ of the sequential circuit $\widetilde V(\Sigma,a;\mathcal{D}, \mathcal{D}')$, we consider a partition of a line that separates $\Sigma$ into small fractions $\{L_{r'}\}$. 

We decompose each sequential circuit into small circuits as
\begin{align}
    \widetilde V(\Sigma,a;\mathcal{D}, \mathcal{D}') = \prod_{L_{r'
    }} U_{L_{r'}}~,
\end{align}
where we take an ordered product of non-commuting circuits $U_{L_{r'}}$. Starting with a defect configuration $\mathcal{D},$ the above sequential circuit transforms $\mathcal{D}$ into $\mathcal{D}'$ through intermediate defect configurations, $\mathcal{D}\to \mathcal{D}_1\to\cdots\to\mathcal{D}_n\to \mathcal{D}'$. Each intermediate defect configuration is associated with 0d states $\ket{\psi}_{{\mathcal{D}}_j}$. The defect states are in a tensor product form:
\begin{align}
    \ket{\psi}_{{\mathcal{D}}_j} = \bigotimes_{p} \ket{\psi}_{{\mathcal{D}}_j|_p}~,
\end{align}
where each state $\ket{\psi}_{{\mathcal{D}_j|}_p}$ is supported at a point $p$ where a defect is located.

Now, suppose we apply a circuit $U_{L_{r'}}$ to the state $\ket{\widetilde{\mathcal{D}}_j}$ and evolve it into $\ket{\widetilde{\mathcal{D}}_{j+1}}$.
Each circuit $U_{L_{r'}}$ acts on a pair of points at the boundary $\partial L_{r'}$, and transforms it into a new one. Concretely, it acts on the Hilbert space within $L_{r'}$ by
\begin{small}
\begin{align}
    &U_{L_{r'}} [\ket{\psi}_{\mathcal{D}_{j}\cap  L_{r'}}\otimes \ket{0}] =\cr  &\exp(i\theta(L_{r'}, a; \mathcal{D}_j|_{L_{r'}},\mathcal{D}_{j+1}|_{L_{r'}})) [\ket{\psi}_{\mathcal{D}_{j+1}\cap  L_{r'}}\otimes \ket{0}]~,
\end{align}
    \end{small}
where $\ket{0}$ denotes the trivial product state of $L_{r'}$ that complements the defect state at the ends.
Here, once we fix the choice of each point state $\ket{\psi}_{{\mathcal{D}}_j|_p}$, $U_{L_{r'}}$ acts by the phase $\theta(L_{r'},a; \mathcal{D}_j|_{L_{r'}},\mathcal{D}_{j+1}|_{L_{r'}})$ that only depends on the local defect configurations at the ends of the line $L_{r'}$.

Due to the condition~\eqref{eq:decompose V}, a sequential circuit $ \widetilde V(\Sigma,a;\mathcal{D}, \mathcal{D}')$ can be decomposed into a form
\begin{align}
    \widetilde V(\Sigma,a;\mathcal{D}, \mathcal{D}') = U'U_{r}(a, (\mathcal{D},\mathcal{D}')|_{D_r})U''~,
\end{align}
such that $U_r$ is supported at an $r$-neighborhood of a generic point (i.e., an interval with length $2r$), and $U', U''$ are away from the point with the distance $r-\eps$.

When $r'$ is taken to satisfy $r'<2r$, 
the phase factor $\theta(L_{r'}, a; \mathcal{D}_j|_{L_{r'}},\mathcal{D}_{j+1}|_{L_{r'}})$ is entirely determined by the circuit $U_{r}(a, (\mathcal{D},\mathcal{D}')|_{D_r})$ within the $r$-neighborhood that contains $L_{r'}.$ In the whole Berry phase, this circuit $U_r(a, (\mathcal{D},\mathcal{D}')|_{D_r})$, accordingly the phase factor $\theta(L_{r'}, a; \mathcal{D}_j|_{L_{r'}},\mathcal{D}_{j+1}|_{L_{r'}})$, appears once in each sequential circuit $\widetilde V(\Sigma,a;\mathcal{D}_\alpha,\mathcal{D}_\beta)$ such that $(\mathcal{D},\mathcal{D}')|_{D_r} =(\mathcal{D}_\alpha,\mathcal{D}_\beta)|_{D_r}$, and its inverse appears in $\widetilde V(\Sigma,a^*;\mathcal{D}_\beta,\mathcal{D}_\alpha)$ such that $(\mathcal{D},\mathcal{D}')|_{D_r} =(\mathcal{D}_\beta,\mathcal{D}_\alpha)|_{D_r}$.

Therefore, once we break each phase factor $\theta(\Sigma,a;\mathcal{D},\mathcal{D}')$ into the sum of local phase factors $\theta(L_{r'}, a;\mathcal{D}_j|_{L_{r'}},\mathcal{D}_{j+1}|_{L_{r'}})$, each local phase factor $\theta(L_{r'}, a;\mathcal{D}_j|_{L_{r'}},\mathcal{D}_{j+1}|_{L_{r'}})$ and its inverse sum up to
\begin{align}
    \left(\sum_{\substack{\mathcal{D},\mathcal{D}' \\ (\mathcal{D},\mathcal{D}')|_{D_{r}} = (\mathcal{D}_j, \mathcal{D}_{j+1})|_{D_r}}}(\epsilon(a;\mathcal{D}, \mathcal{D}')-\epsilon(a^*;\mathcal{D}', \mathcal{D})) \right)&\times \cr \theta(L_{r'}, a;\mathcal{D}_j|_{L_{r'}},\mathcal{D}_{j+1}|_{L_{r'}}) &= 0,
\end{align}
where we used Eq.~\eqref{eq:epsilon locality constraint}. This implies that each local Berry phase cancels out. Therefore, the whole Berry phase must vanish on an SRE state.

\subsubsection{When defects are lines}

\begin{figure*}[t]
    \centering
    \includegraphics[width=0.8\linewidth]{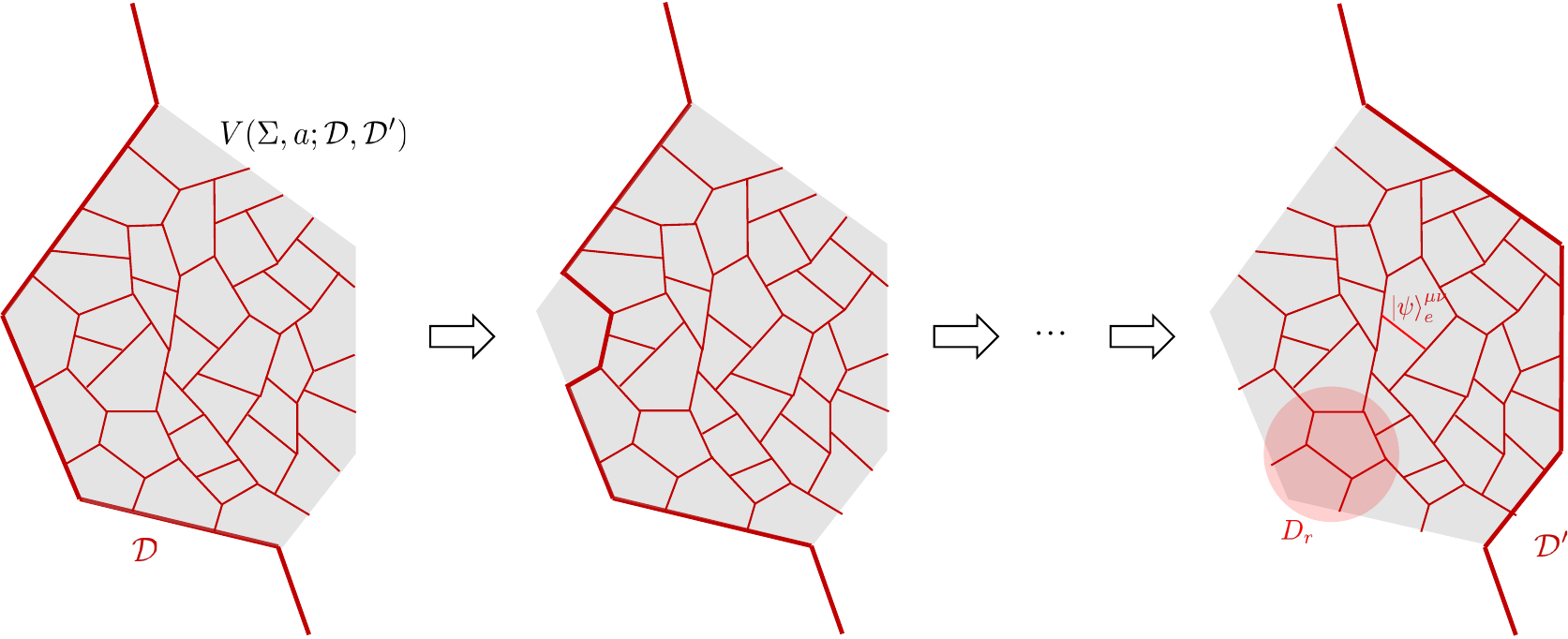}
    \caption{We break a sequential circuit $V(\Sigma,a;\mathcal{D},\mathcal{D}')$ into small fractions $\{L_{r'}\}$, so that each fraction is contained in a disk $D_r$ of radius $r$ (red disk). A process of moving a defect breaks into $\mathcal{D}\to \mathcal{D}_1\to...\mathcal{D}_n\to \mathcal{D}'$ via intermediate configurations. Each edge $e$ of the fine network supports an MPS state $\ket{\psi}_{e}^{\mu\nu}$, which constitutes a 1d state localized along the defect. Each local phase factor $\theta(L_{r'}, a;\mathcal{D}_j|_{L_{r'}},\mathcal{D}_{j+1}|_{L_{r'}})$ cancels out due to the locality constraints on integer coefficients $\epsilon$. }
    \label{fig:breakcircuit}
\end{figure*}

Let us now consider the case when the defects are 1d objects; the defect state $\ket{\widetilde{\mathcal{D}}}$ has a 1d localized state along the defect network,
\begin{align}
\ket{\widetilde{\mathcal{D}}} = \ket{\psi}_{\mathcal{D}}\otimes \ket{0}~,
\end{align}
where $\ket{\psi}_{\mathcal{D}}$ is regarded as a 1d gapped state localized at the defect network, which admits a matrix product state (MPS) representation.

For each 2d operator support $\Sigma$ of the sequential circuit $\widetilde V(\Sigma,a;\mathcal{D}, \mathcal{D}')$, we consider a fine mesh of 1d network within $\Sigma$ that separates $\Sigma$ into small 2d regions $\{L_{r'}\}$. Each region $L_{r'}$ is contained in a ball with radius $r'<r$, See Fig.~\ref{fig:breakcircuit}.

We decompose each sequential circuit into small circuits as
\begin{align}
    \widetilde V(\Sigma,a;\mathcal{D}, \mathcal{D}') = \prod_{L_{r'
    }} U_{L_{r'}}~,
\end{align}
where we take an ordered product of non-commuting circuits $U_{L_{r'}}$. Starting with a defect network configuration $\mathcal{D},$ the above sequential circuit transforms $\mathcal{D}$ into $\mathcal{D}'$ through intermediate defect configurations, $\mathcal{D}\to \mathcal{D}_1\to\cdots\to\mathcal{D}_n\to \mathcal{D}'$. Each intermediate defect configuration is associated with an MPS $\ket{\psi}_{{\mathcal{D}}_j}$. To describe these intermediate MPS, it is convenient to introduce a ``patchwork'' of MPS representations: we consider a fraction of an MPS supported at each edge of the network $\ket{\psi}_e^{\mu\nu}$, where $\mu,\nu$ are indices of bond Hilbert spaces at the boundary of an edge $e$. 

Each MPS $\ket{\psi}_{{\mathcal{D}}_j}$ is then represented by a ``patchwork'' of MPS patches $\ket{\psi}_e^{\mu\nu}$ on each network edge within the defect $\mathcal{D}_j$:
\begin{align}
    \ket{\psi}_{{\mathcal{D}}_j} = \mathrm{Tr}(...\ket{\psi}_{e_k}^{\mu\nu}\ket{\psi}_{e_{k+1}}^{\nu\lambda}...)
\end{align}
which combines MPS patches at the edges $\{e_k\}$ supported on the defect position of $\mathcal{D}_j$ \footnote{For simplicity, the above expression assumed that the defect configuration $\mathcal{D}_j$ does not have junctions, and is equivalent to a single closed loop.}.

Now, suppose we apply a circuit $U_{L_{r'}}$ to the state $\ket{\widetilde{\mathcal{D}}_j}$ to evolve it into $\ket{\widetilde{\mathcal{D}}_{j+1}}$.
Each circuit $U_{L_{r'}}$ acts on a fraction of MPS patchworks at the boundary $\partial L_{r'}$, and transforms it into a new one. Concretely, it acts on the Hilbert space within $L_{r'}$ by
\begin{small}
\begin{align}
    &U_{L_{r'}} [\ket{\psi}^{\mu\nu}_{\mathcal{D}_{j}\cap  L_{r'}}\otimes \ket{0}] =\cr  &\exp(i\theta(L_{r'}, a; \mathcal{D}_j|_{L_{r'}},\mathcal{D}_{j+1}|_{L_{r'}})) [\ket{\psi}^{\mu\nu}_{\mathcal{D}_{j+1}\cap  L_{r'}}\otimes \ket{0}]~,
\end{align}
\end{small}
where $\ket{0}$ denotes the trivial product state of $L_{r'}$ that complements the defect state.
Here, once we fix the choice of MPS patches, $U_{L_{r'}}$ acts by the phase $\theta(L_{r'}, a;\mathcal{D}_j|_{L_{r'}},\mathcal{D}_{j+1}|_{L_{r'}})$ that only depends on the local defect configurations in the vicinity of the region $L_{r'}$.

Due to the condition~\eqref{eq:decompose V}, a sequential circuit $\widetilde V(\Sigma,a;\mathcal{D}, \mathcal{D}')$ can be decomposed into a form
\begin{align}
   \widetilde V(\Sigma,a;\mathcal{D}, \mathcal{D}') = U'U_{r}(a, (\mathcal{D},\mathcal{D}')|_{D_r})U''~,
\end{align}
such that $U_r$ is supported at an $r$-neighborhood of a generic point, and $U', U''$ are away from the point with the distance $r-\eps$.

Since $L_{r'}$ is supported within a ball of size $r'<r$, 
the phase factor $\theta(L_{r'}, a; \mathcal{D}_j|_{L_{r'}},\mathcal{D}_{j+1}|_{L_{r'}})$ is entirely determined by the circuit $U_{r}(a, (\mathcal{D},\mathcal{D}')|_{D_r})$ within the $r$-neighborhood that contains $L_{r'}.$ In the whole Berry phase, this circuit $U_r(a, (\mathcal{D},\mathcal{D}')|_{D_r})$, accordingly the phase factor $\theta(L_{r'}, a; \mathcal{D}_j|_{L_{r'}},\mathcal{D}_{j+1}|_{L_{r'}})$, appears once in each sequential circuit $\widetilde V(\Sigma,a;\mathcal{D}_\alpha,\mathcal{D}_\beta)$ such that $(\mathcal{D},\mathcal{D}')|_{D_r} =(\mathcal{D}_\alpha,\mathcal{D}_\beta)|_{D_r}$, and its inverse appears in $\widetilde V(\Sigma,a;\mathcal{D}_\beta,\mathcal{D}_\alpha)$ such that $(\mathcal{D},\mathcal{D}')|_{D_r} =(\mathcal{D}_\beta,\mathcal{D}_\alpha)|_{D_r}$. 

Therefore, once we break each phase factor $\theta(\Sigma,a;\mathcal{D},\mathcal{D}')$ into the sum of local phase factors $\theta(L_{r'}, a;\mathcal{D}_j|_{L_{r'}},\mathcal{D}_{j+1}|_{L_{r'}})$, each local phase factor $\theta(L_{r'}, a;\mathcal{D}_j|_{L_{r'}},\mathcal{D}_{j+1}|_{L_{r'}})$ and its inverse sum up to
\begin{align}
    \left(\sum_{\substack{\mathcal{D},\mathcal{D}' \\ (\mathcal{D},\mathcal{D}')|_{D_{r}} = (\mathcal{D}_j, \mathcal{D}_{j+1})|_{D_r}}}(\epsilon(a;\mathcal{D}, \mathcal{D}')-\epsilon(a^*;\mathcal{D}', \mathcal{D})) \right)&\times \cr \theta(L_{r'}, a;\mathcal{D}_j|_{L_{r'}},\mathcal{D}_{j+1}|_{L_{r'}}) &= 0,
\end{align}
where we used Eq.~\eqref{eq:epsilon locality constraint}. This implies that each local Berry phase cancels out. Therefore, the whole Berry phase must vanish on an SRE state.

\section{Non-Abelian fermionic loop in (3+1)D}
\label{sec:floop}
In this section, we apply the framework of the invariants to loop excitations in three spatial dimensions.
We identify a loop excitation in (3+1)D $\mathbb{D}_4$ topological order called a non-Abelian fermionic loop. 

To understand a non-Abelian fermionic loop in continuum QFT, we first recall that the $\Z_2$ gauge theory has a single fermionic loop; consider a (3+1)D $\Z_2$ gauge theory with the action
\begin{align}
    \exp\left(\pi i \int bda\right)
\end{align}
with $a,b$ 1, 2-form $\Z_2$ gauge fields. The fermionic loop excitation corresponds to a topological surface operator
\begin{align}
    S_f(\Sigma) = \exp\left(\pi i \int_\Sigma b+a\cup a\right)~,
    \label{eq:abelian f loop}
\end{align}
which has an 't Hooft anomaly and exhibits the fermionic loop self-statistics \cite{hsin2025higherformanomaliesimplyintrinsic}.
The non-Abelian fermionic loop of the $\mathbb{D}_4$ gauge theory is then understood by expressing the gauge group $\mathbb{D}_4=(\Z^A_2\times\Z^B_2)\rtimes\Z^C_2$, where $\Z_2^C$ acts by swapping generators of $\Z_2^A, \Z_2^B$. One can write a surface operator $S_f(\Sigma)$ for each of the $\Z_2^A$ and $\Z_2^B$ gauge fields, but neither of them is invariant under $\Z_2^C$ gauge transformation. However, there is a gauge invariant superposition of fermionic loops for $\Z_2^A, \Z_2^B$ that leads to a non-invertible operator, a non-Abelian fermionic loop. This is a coset non-invertible symmetry obtained by gauging a non-normal subgroup symmetry~\cite{Hsin2024coset, Hsin2026coset}; see e.g., \cite{Thorngren:2021yso,Chatterjee:2024ych, Heidenreich:2021xpr, Arias-Tamargo:2022nlf, Antinucci2022continuous, Nguyen_2021, Bhardwaj_2023, schafernameki2023ictp, jacobson2024gaugingclattice} for other examples of coset symmetries.

\subsection{$\D_4$ topological order in (3+1)D}

Here we find a non-Abelian fermionic loop operator in the lattice Hamiltonian model of $\D_4$ gauge theory on a 3D cubic lattice. This is an analog of the quantum double model on a 2D square lattice; the local Hilbert space is defined on each edge and is spanned by the basis states $\{\ket{g}\}$ labeled by $g\in \D_4$. For later convenience, we also introduce an alternative description of the same topological order using qudits. 

\subsubsection{Lattice Hamiltonian}
\begin{figure}
    \centering
    \includegraphics[width=1.0\linewidth]{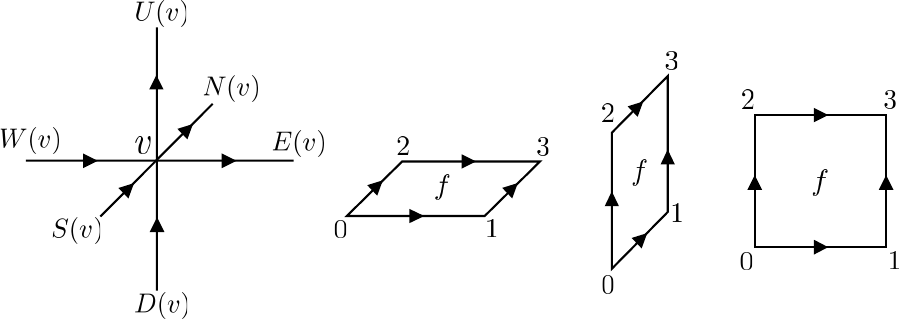}
    \caption{Labels and orientations of a vertex and faces on the cubic lattice.}
    \label{fig:cube}
\end{figure}

The Hamiltonian on a 3D cubic lattice is defined as
\begin{align}
    H = -\sum_{v}A_v-\sum_{f}B_f
\end{align}
where the vertex term $A_v$ and the face term $B_f$ are given by
\begin{align} 
A_v &\coloneq \frac{1}{|\D_4|}\sum_{g\in \D_4}\overrightarrow{X}_{E(v)}^g\lX_{N(v)}^g\lX_{U(v)}^g\rX_{W(v)}^g\rX_{S(v)}^g\rX_{D(v)}^g,\\
B_f &\coloneq \delta_{g_{01}g_{13}g_{23}^{-1}g_{02}^{-1},e},
\end{align}
 where $e$ is an identity element in a group, and an orientation of an edge is shown in Fig.~\ref{fig:cube}.
We also defined the operator acting on an edge $e$ as
\begin{align}
    \lX^g\ket{h} =\ket{gh}~,\quad  \rX^g\ket{h} =\ket{hg^{-1}}~,
\end{align}
These vertex and face operators are commutative, and the ground state of the model satisfies the Gauss law constraint $A_v = 1$ as well as the flatness conditions of the gauge fields $B_f = 1$.

For our purpose of describing the non-Abelian fermionic loop, it is convenient to represent the gauge group as $\D_4=(\Z_2\times\Z_2)\rtimes\Z_2$; an element $g\in \D_4$ is labeled by a pair $((g_A, g_B), g_C)$ with $(g_A, g_B) \in \Z^{A}_2\times\Z^{B}_2, g_C\in \Z^{C}_2$, where $\Z^{C}_2$ acts on $\Z^{A}_2\times\Z^{B}_2$ by an automorphism swapping the $\Z_2^{A}, \Z_2^{B}$. The multiplication law is explicitly written as
\begin{align}
&((g_A, g_B), g_C)\cdot  ((g_A^\prime, g_B^\prime), g_C^\prime) \nonumber\\&=
\begin{cases}
((g_A + g_A^\prime,\, g_B + g_B^\prime),\; g_C + g_C^\prime), &  \text{for}\ g_C = 0, \\[1mm]
((g_A + g_B^\prime,\, g_B + g_A^\prime),\; g_C + g_C^\prime), &  \text{for}\ g_C = 1,
\end{cases}    
\end{align}
where all sums are mod $2$. Due to the identification $\D_4=(\Z_2\times\Z_2)\rtimes\Z_2$, each onsite Hilbert space $\{\ket{g}_e\}$ with $g\in\D_4$ is identified as three qubits, $\{X_e^{i},Z_e^{i}\}$ with $i=A,B,C$.

One can then represent the Hamiltonians using the qubit operators: for instance, $\lX^g_{e}$ with $g=((g_A,g_B),g_C)$ is represented as
\begin{align}
    \lX^g_{e}= \left(X_e^{A}\right)^{g_A}\left(X_e^{B}\right)^{g_B}\left(X^{C}_e \text{SWAP}_e^{{(AB)}}\right)^{g_C},
\end{align}
where $\mathrm{SWAP}_e^{(ij)}$ is a swap gate between labels $i$ and $j$ on the edge $e$. $\rX^g_{e}$ is given by
\begin{small}
\begin{align}
         \rX^g_{e} =&\left[\text{CSWAP}_{e,e}^{(C),(AB)}\right]\left(X_e^{A}\right)^{g_A}\left(X_e^{B}\right)^{g_B}\left[\text{CSWAP}_{e,e}^{(C),(AB)}\right]^{-1}\cr &\times \left(X_e^{C}\right)^{g_C}~,
\end{align}
\end{small}
where $\text{CSWAP}_{e_1,e_2}^{(C),(AB)}$ is a controlled SWAP gate $\text{CSWAP}^{(C),(AB)} \coloneq I_{e_2}\otimes \ketbra{0_C}_{e_1}+\mathrm{SWAP}^{(AB)}_{e_2}\otimes \ketbra{1_C}_{e_1}$.
The Hamiltonian is then expressed as
\begin{small}
\begin{align}
    A_v = &\prod_{\alpha=A,B}\frac{1}{2}\left(1+\prod_{e= U(v),E(v),N(v)}X^{\alpha}_e\prod_{e= W(v),S(v)}X^{^{c(e,e)}(\alpha)}_e\right)\nonumber\\
    &\times \frac{1}{2}\left(1+\prod_{e= U(v),E(v),N(v)}\mathrm{SWAP}^{(AB)}_eX^{C}_e\prod_{e= W(v),S(v)}X^{C}_e\right),\\
    B_f = &\prod_{\alpha=A,B}\frac{1}{2}\left(1+Z_{e_{01}}^{\alpha}Z^{^{c(01,13)}(\alpha)}_{e_{13}}Z^{\alpha}_{e_{02}}Z^{^{c(02,23)}(\alpha)}_{e_{23}}\right)\nonumber\\
    &\times \frac{1}{2}\left(1+Z^{C}_{e_{01}}Z^{C}_{e_{13}}Z^{C}_{e_{02}}Z^{C}_{e_{23}}\right),
\end{align}
\end{small}
where $^{c(e,e^\prime)}(\alpha)$ represents the controlled SWAP for labels of gauge fields $A,B$ on the edge $e^\prime$, i.e., when $Z^C_e=1$, $^{c(e,e^\prime)}(\alpha)=\alpha$ and when $Z^C_e=-1$, $^{c(e,e^\prime)}(\alpha)=\mathrm{SWAP}(\alpha)$, where we define $\mathrm{SWAP}(A)=B,\mathrm{SWAP}(B)=A$.
We introduce $\ket{\text{GS}}$ as a ground state of $A_v$ and $B_f$ for all vertices and faces.

\subsubsection{Qudit representation of the model}\label{subsubsec:qudit_rep}

We introduce the qudit description of the $\D_4$ gauge theory. This can be achieved by considering the $(\Z_4^A\times \Z^B_4)\rtimes \Z^C_2$ gauge theory with the condensation of the magnetic flux $(m^i)^2$ in each $\Z_4$ toric code part. An explicit construction is as follows. We denote $\Z_4$ Pauli operators as $\widetilde{Z}$ and $\widetilde{X}$, satisfying $\widetilde{Z}\widetilde{X} = i \widetilde{X}\widetilde{Z}$. We now define the Hamiltonian of $\D_4$ gauge theory in (3+1)D in terms of qudits as
\begin{align}
    \widetilde{H}_{\D_4}\coloneq -\sum_{\alpha=A,B}\sum_e \left(\widetilde{X}^{\alpha}_e\right)^2-\sum_v\left(\widetilde{A}_v + \widetilde{A}_v^\dagger\right) -\sum_f \widetilde{B}_f,
\end{align}
where $\widetilde{A}_v$ is the Gauss law constraint for the $\D_4$ gauge theory with $X$ replaced by $\widetilde{X}$, and $\widetilde{B}_f$ for a face $f=(0123)$ is defined by
\begin{scriptsize}
\begin{align}
    \widetilde{B}_f\coloneq &\prod_{\alpha=A,B}\frac{1}{2}\left\{1+\left(\widetilde{Z}_{e_{01}}^{\alpha}\right)^2\left(\widetilde{Z}^{^{c(01,13)}(\alpha)}_{e_{13}}\right)^2\left(\widetilde{Z}^{\alpha}_{e_{02}}\right)^2\left(\widetilde{Z}^{^{c(02,23)}(\alpha)}_{e_{23}}\right)^2\right\}\nonumber\\
    &\times \frac{1}{2}\left(1+Z^{C}_{e_{01}}Z^{C}_{e_{13}}Z^{C}_{e_{02}}Z^{C}_{e_{23}}\right).
\end{align}
\end{scriptsize}
The squared terms $\left(\widetilde{X}^{(i)}_e\right)^2$ energetically restrict the theory to an effective
$\left(\Z_2^A\times\Z_2^B\right)\rtimes \Z_2^C$ gauge theory.

\subsection{Non-Abelian fermionic loops in $\mathbb{D}_4$ gauge theory}

We observe that the \(\D_4 = (\Z_2^{A} \times \Z_2^{B}) \rtimes \Z_2^{C}\) gauge theory can be obtained by starting with two copies of a \(\Z_2\) gauge theory (\(\Z_2^{A} \times \Z_2^{B}\) gauge theory) and then gauging the \(\Z_2\) symmetry that swaps the two copies. Because each copy of the \(\Z_2\) gauge theory contains a fermionic loop excitation, gauging the SWAP symmetry produces a non-Abelian fermionic loop in the resulting \(\D_4\) gauge theory. Concretely, this loop arises as a gauge-invariant superposition of the fermionic loops in the two layers.
In this section, we construct this non-Abelian fermionic loop in the \(\D_4\) gauge theory explicitly from this perspective.

\subsubsection{Warm-up: fermionic loop in $(3+1)$D $\Z_2$ gauge theory}

Let us first recall the Abelian fermionic loop in $(3+1)$D $\Z_2$ gauge theory on the cubic lattice~\cite{hsin2025higherformanomaliesimplyintrinsic}.
Let us introduce the Hamiltonian
\begin{align}
    H_{\Z_2} = -\sum_{v}\prod_{\partial e\ni v}X_e - \sum_{f}\prod_{e\in \partial f}Z_e
\end{align}
where a $\Z_2$ qubit lives on each edge, and $X_e$ and $Z_e$ are Pauli operators.

\begin{figure}[t]
    \centering
    \includegraphics[width=1.0\linewidth]{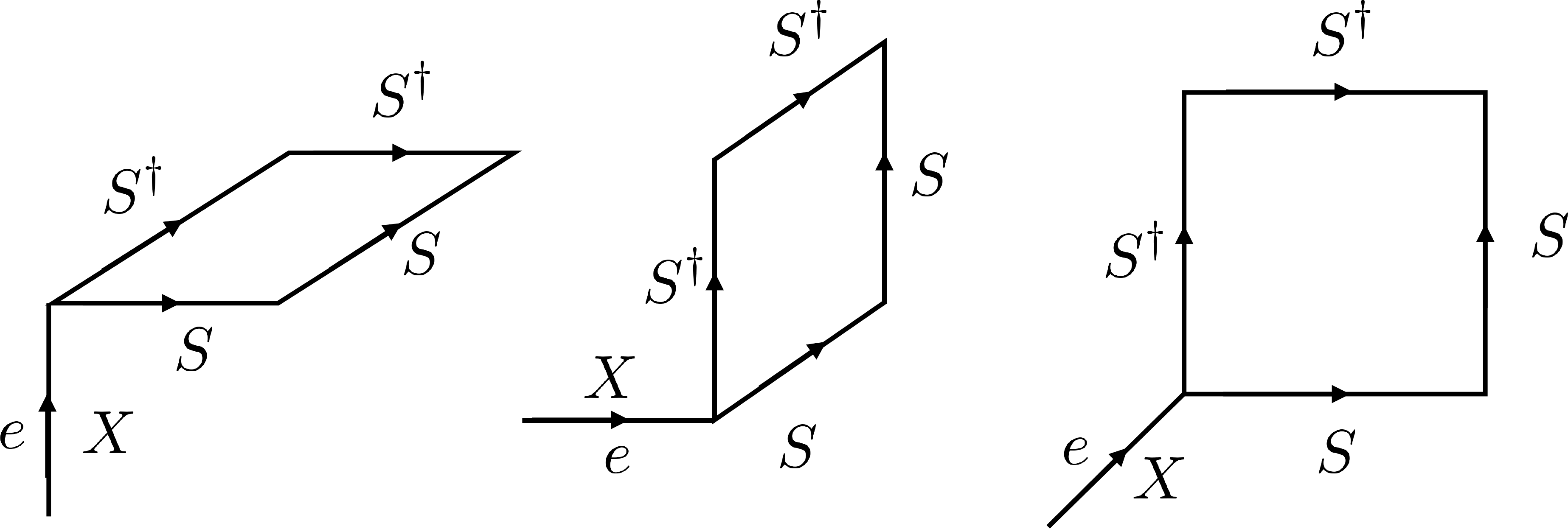}
    \caption{Definition of the operator $S_e$.}
    \label{fig:abelian_surface_op}
\end{figure}

\begin{figure}[t]
    \centering
    \includegraphics[width=0.7\linewidth]{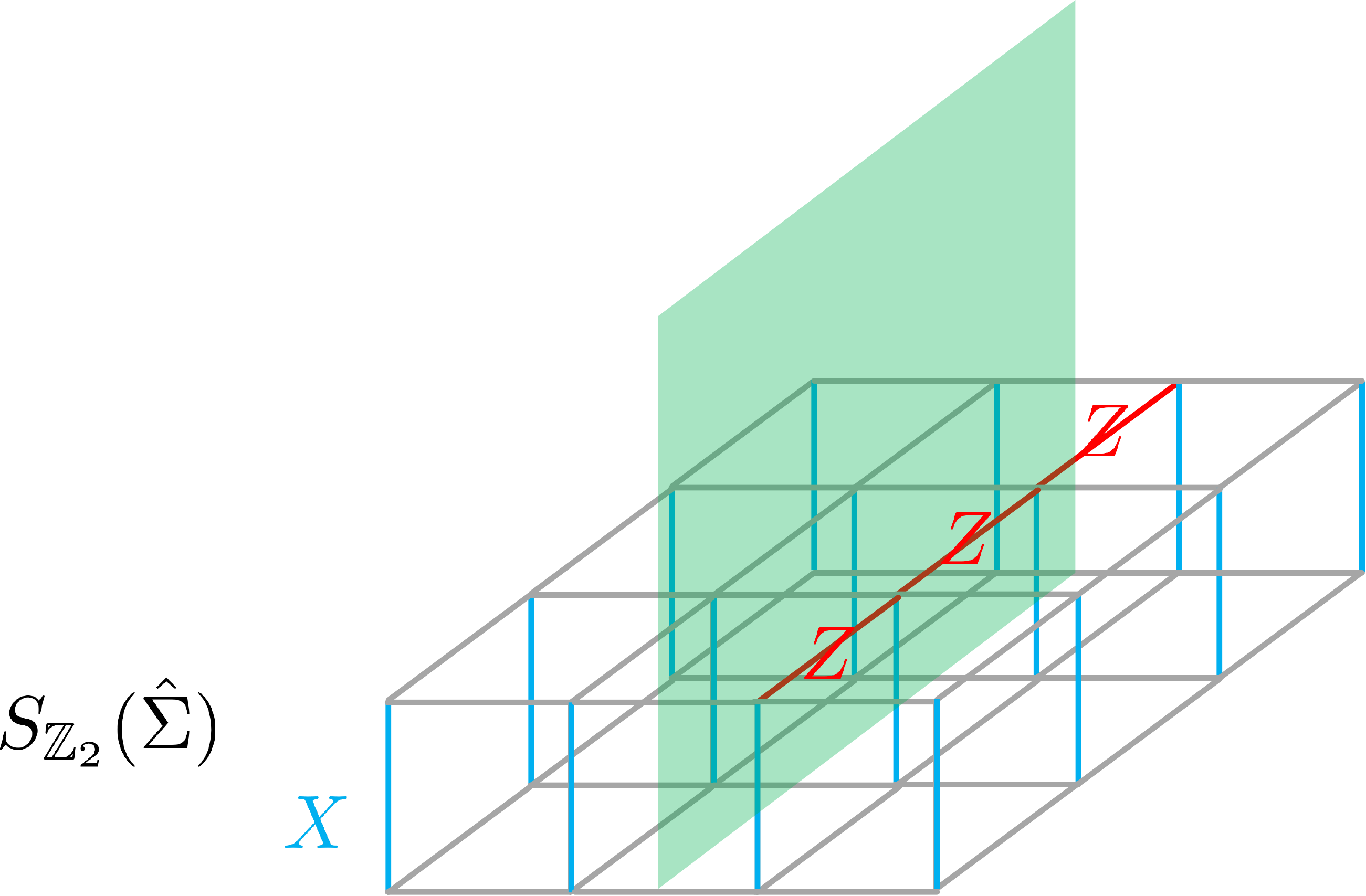}
    \caption{The surface operator $S_{\Z_2}(\hat\Sigma)$ on a Klein bottle. The Pauli $X$ operators are supported on blue edges, while the Pauli $Z$ operators are supported on red ones. The red edges correspond to the intersection between the surface $\Sigma$ and the orientation-reversing domain wall (green sheet).}
    \label{fig:rev_domain_wall}
\end{figure}

The theory has an anomalous $\Z_2$ $1$-form symmetry generated by the surface operator
\begin{align}
    S_{\Z_2}(\hat\Sigma) = \prod_{e\in \hat\Sigma}S_e,
    \label{eq:Se}
\end{align}
where $\hat{\Sigma}$ is a surface of the dual lattice, and $S_e$ is defined in Fig.~\ref{fig:abelian_surface_op} with $S$ gate $S\coloneq \text{diag}(1,i)$. The product is over the edges intersecting $\hat\Sigma$.
When $\hat\Sigma$ is oriented, the contributions of $S, S^\dagger$ operators cancel out and $S_{\Z_2}(\hat\Sigma)$ simply gives a magnetic surface operator of the $\Z_2$ toric code. 
Meanwhile,
when $\hat\Sigma$ is non-orientable, $S_{\Z_2}(\hat\Sigma)$ differs from the pure magnetic operator by the additional electric $S$ operators; the electric operators form a string of Pauli $Z$ operators along the orientation reversing defect, see Fig.~\ref{fig:rev_domain_wall}.
The orientation reversing defect supports a string of $XZ$ operators, and is therefore expected to exhibit fermionic statistics.

\begin{figure}[t]
    \centering
    \includegraphics[width=0.75\linewidth]{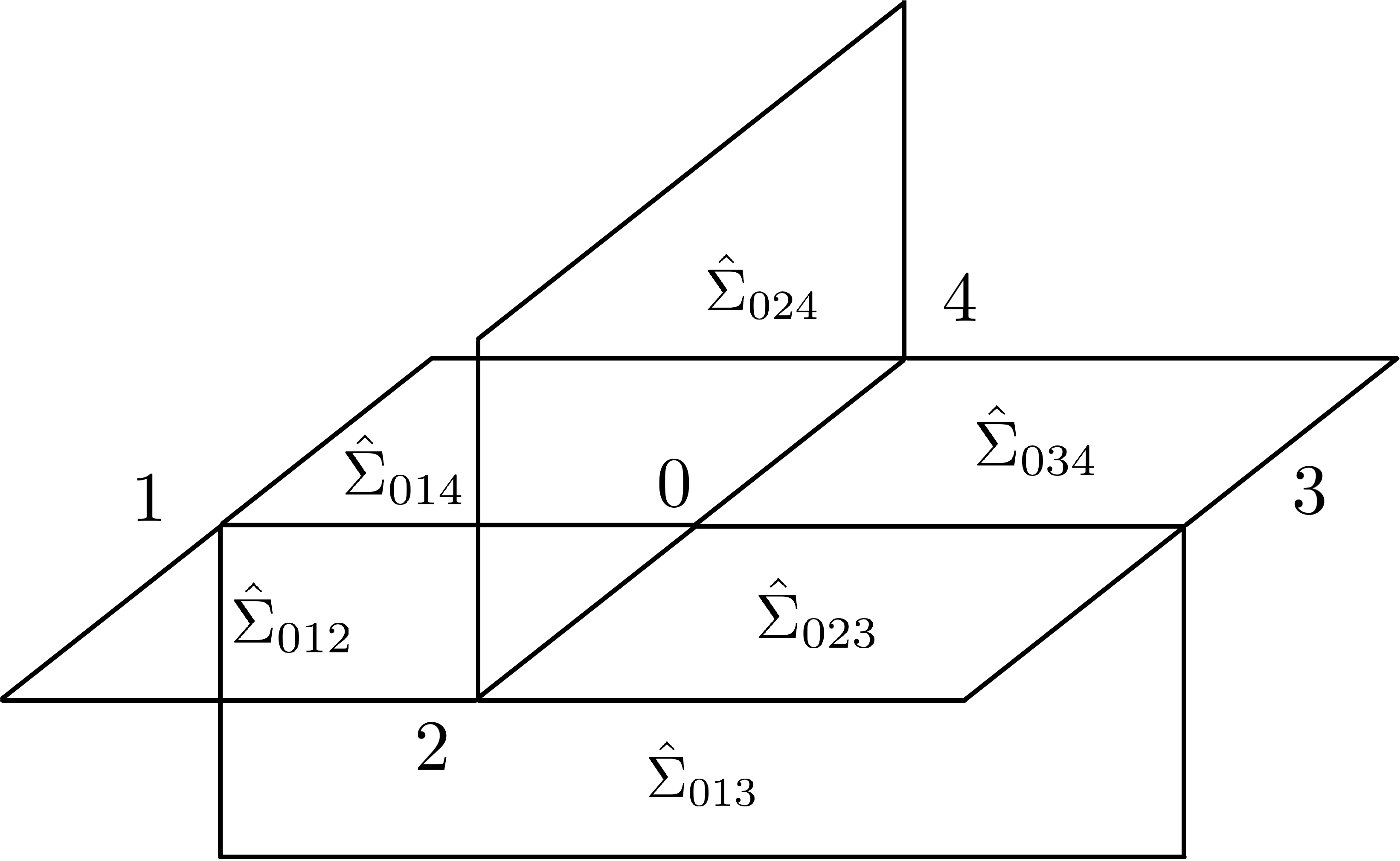}
    \caption{Open surfaces $\hat\Sigma_{0ij}$ defined on the dual cubic lattice.}
    \label{fig:surfaces}
\end{figure}

\begin{figure*}[t]
    \centering
    \includegraphics[width=1.0\linewidth]{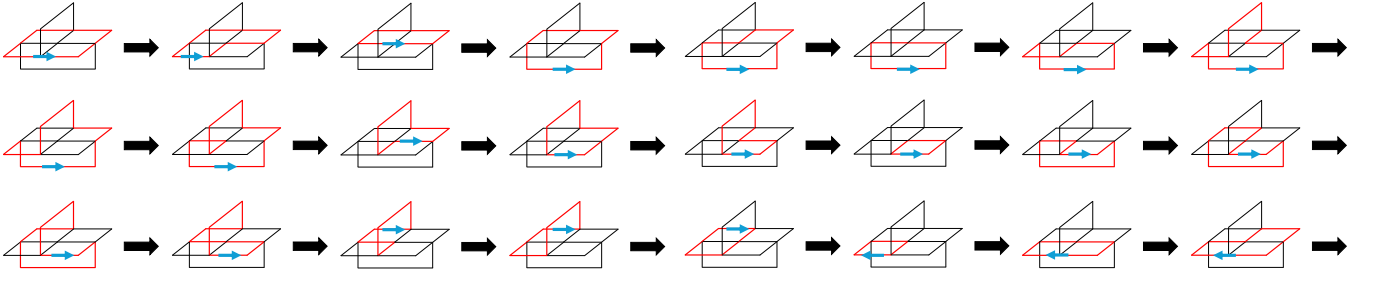}
    \caption{24-unitary sequence to define the invariant. Red lines denote the configuration of the loop excitations, and blue arrows show the orientation of the configurations. Applying this unitary sequence returns the configuration to the initial one, but with the orientation reversed.}
    \label{fig:24-sequence}
\end{figure*}

The self-statistics of the (Abelian) fermionic loop is characterized by the following $\Z_2$ invariant of finite-depth unitaries~\cite{Kobayashi2026generalized} for the surface operator:
\begin{small}
\begin{align}
    &U_\Theta = \cr &S_{\Z_2,014}S_{\Z_2,034}S_{\Z_2,023}S_{\Z_2,014}^{-1}S_{\Z_2,024}^{-1}S_{\Z_2,012}S_{\Z_2,023}^{-1}S_{\Z_2,013}^{-1}\times\nonumber\\
    & S_{\Z_2,024}S_{\Z_2,014}S_{\Z_2,013}S_{\Z_2,024}^{-1}S_{\Z_2,034}^{-1}S_{\Z_2,023}S_{\Z_2,013}^{-1}S_{\Z_2,012}^{-1}\times\nonumber\\
    &S_{\Z_2,034}S_{\Z_2,024}S_{\Z_2,012}S_{\Z_2,034}^{-1}S_{\Z_2,014}^{-1}S_{\Z_2,013}S_{\Z_2,012}^{-1}S_{\Z_2,023}^{-1}~,
\end{align}
\end{small}
where $S_{\Z_2,0ij}$ is the symmetry operator on an open surface $\hat\Sigma_{0ij}$ as shown in Figs.~\ref{fig:surfaces} and~\ref{fig:24-sequence}. This 24-unitary sequence flips the orientation of the loop excitation, and also satisfies the required conditions in Eqs.~\eqref{eq:redefining constraint} and~\eqref{eq:epsilon locality constraint}, therefore defines an invariant.
This invariant becomes nontrivial on the fermionic loop operators $S_{\mathbb{Z}_2}$~\cite{hsin2025higherformanomaliesimplyintrinsic}; $U_\Theta = -1$, showing that the 1-form $\Z_2$ symmetry is anomalous. The nontrivial self-statistics captured by the value of $U_\Theta=-1$ correspond to the fermionic loop statistics.

\subsubsection{Non-Abelian fermionic loop in (3+1)D $\D_4$ gauge theory}

Let us now consider a non-Abelian fermionic loop of $\D_4$ gauge theory.
We define the following surface operator on a closed surface $\hat\Sigma$ on the dual cubic lattice:
\begin{align}
    S_{\D_4}(\hat\Sigma)&\coloneq W_S(\Sigma)\cdot \frac{1}{2}\sum_{\alpha=A,B}T_X^{\alpha}(\hat\Sigma)\cdot\cr &\prod_{\gamma\in H_1(\Sigma,\Z_2)}\frac{1+\prod_{e\in \gamma}Z^{C}_{e}}{2}\prod_{f\in \Sigma}\frac{1+\prod_{e\in \partial f}Z^{C}_{e}}{2}~,\\
    \label{eq:SD8}
    T_X^{\alpha}(\hat\Sigma)&\coloneq \prod_{e\in\hat\Sigma}X^{^{c_{\gamma}(e_0,e)}(\alpha)}_{e}~,\\
    W_S(\Sigma)&\coloneq \prod_{\alpha=A,B}\prod_{f=(0123)\in\Sigma}S^{\alpha}_{e_{01}} S^{\alpha}_{e_{13}}{S^{\alpha\dagger}_{e_{02}}} {S^{\alpha\dagger}_{e_{23}}}~,
\end{align} 
where $\Sigma$ is a surface of the original cubic lattice obtained by translating $\hat\Sigma$ by $(1/2,1/2,1/2)$. In the above expression of the operator $T_X^{\alpha}(\hat\Sigma)$, we fix a choice of an edge $e_0\in\hat\Sigma$ that intersects $\hat{\Sigma}$, and define the label $^{c_\gamma(e_0,e)}(\alpha)\in\{A,B\}$ with $\alpha\in\{A,B\}$ as follows. We denote by ${\Sigma}^\prime$ a surface of the original cubic lattice obtained by translating $\hat\Sigma$ by $(0,0,-1/2)$. Suppose $e_0=\langle\overrightarrow{v_0 v'_0}\rangle$, $e=\langle\overrightarrow{v v'}\rangle$, so that vertices $v'_0, v'$ are supported at $\Sigma^\prime$. We then consider a path $\gamma(v'_0,v')$ in $\Sigma$ connecting $v'_0,v'$. We then define path-wise controlled SWAP for labels
\begin{align}
    &^{c_{\gamma}(e_0,e)}(\alpha)\coloneq
    \begin{cases}
     \alpha, &\text{when}\ \prod_{e\in \gamma(v'_0,v')}Z^{C}_{e} = 1~, \nonumber\\
     \text{SWAP}(\alpha), &\text{when}\ \prod_{e\in \gamma(v'_0,v')}Z^{C}_{e} = -1~,
     \end{cases}
\end{align}
where we define SWAP$(A)=B,$ SWAP$(B)=A$. Due to the projectors in the last two terms of $S_{\D_4}(\hat\Sigma)$ in Eq.~\eqref{eq:SD8}, the operator  $S_{\D_4}(\hat\Sigma)$ is independent of the choice of the path $\gamma(v'_0,v')$. One can also verify that $S_{\D_4}(\hat\Sigma)$ is independent of the choice of the ``base'' edge $e_0$; changing $e_0$ at most has the effect of swapping the label $(\alpha)$ in $T_X^{\alpha}$, and summing over $\alpha$ in $S_{\D_4}(\hat\Sigma)$ makes its expression invariant.

This surface operator has the non-invertible fusion rule:
\begin{align}
    & S_{\D_4}(\hat\Sigma)\times S_{\D_4}(\hat\Sigma)= \nonumber \\
    & \left(1+X^{AB}(\hat\Sigma)\right)\prod_{\gamma\in H_1(\Sigma,\Z_2)}\frac{1+\prod_{e\in \gamma}Z^{C}_{e}}{2}\prod_{f\in \Sigma}\frac{1+\prod_{e\in \partial f}Z^{C}_{e}}{2}
\end{align}
with the magnetic surface operator
\begin{align}
    X^{AB}(\hat\Sigma) = \prod_{e\in\hat\Sigma} X^{A}_eX^{B}_e~.
\end{align}
Similarly to the Abelian case, when $\hat\Sigma$ is non-orientable, $S_{\D_4}(\hat\Sigma)$ supports the string of $Z^{A}Z^{B}$ at its orientation-reversing defect. 

In the $\Z_4$ qudit representation, we can similarly define the fermionic loop operator $\widetilde{S}_{\D_4}(\hat\Sigma)$ as 
\begin{align}
    \widetilde{S}_{\D_4}(\hat\Sigma)&\coloneq \widetilde{W}_S(\Sigma)\cdot \frac{1}{2}\sum_{\alpha=A,B}\widetilde{T}_X^{\alpha}(\hat\Sigma)\cdot\cr &\prod_{\gamma\in H_1(\Sigma,\Z_2)}\frac{1+\prod_{e\in \gamma}Z^{C}_{e}}{2}\prod_{f\in \Sigma}\frac{1+\prod_{e\in \partial f}Z^{C}_{e}}{2}~,\\
    \label{eq:SD8_qudit}
    \widetilde{T}_X^{\alpha}(\hat\Sigma)&\coloneq \prod_{e\in\hat\Sigma}\widetilde{X}^{^{c_{\gamma}(e_0,e)}(\alpha)}_{e}~,\\
    \widetilde{W}_S(\Sigma)&\coloneq \prod_{\alpha=A,B}\prod_{f=(0123)\in\Sigma}\widetilde{Z}^{\alpha}_{e_{01}} \widetilde{Z}^{\alpha}_{e_{13}}{\widetilde{Z}^{\alpha\dagger}_{e_{02}}} {\widetilde{Z}^{\alpha\dagger}_{e_{23}}}~.
\end{align}

\subsection{Self-statistics of non-Abelian fermionic loop}

We introduce a local unitary circuit to deform defect states. For later purposes, we consider the two types of defect states:
\begin{align}
    \ket{\mathcal{D}^f_{\hat\Sigma}}&\coloneq S_{\D_4}(\hat\Sigma)\ket{\text{GS}}~,\cr  \quad \ket{\mathcal{D}^b_{\hat\Sigma}}&\coloneq \frac{1}{\sqrt{2}}\left(T^A_X(\hat\Sigma)+T^B_X(\hat\Sigma)\right)\ket{\text{GS}}~,
\end{align}
where $\ket{\text{GS}}$ is a ground state of $(3+1)$D $\D_4$ gauge theory and $\hat\Sigma$ is an open surface. 
The state $\ket{\mathcal{D}^f_{\hat\Sigma}}$ has \textit{fermionic} loop excitations, while $\ket{\mathcal{D}^b_{\hat\Sigma}}$ has standard \textit{bosonic} magnetic flux excitations.

Let us consider a sequential circuit $V^f$ for fermionic loop $\Z_2$ 1-form symmetry to deform the defect configurations from $\mathcal{D}^f_{\hat\Sigma}$ to $\mathcal{D}^f_{\hat\Sigma\cup \hat\Sigma_1}$ in the form of
\begin{align}
    V^f(\hat\Sigma,\hat\Sigma_1) = W_S(\Sigma_1)V^b(\hat\Sigma,\hat\Sigma_1),
    \label{eq:Vf}
\end{align}
where $V^b(\hat\Sigma,\hat\Sigma_1)$ is a sequential unitary supported at $\hat\Sigma_1$ to move non-Abelian bosonic excitation from $\hat\Sigma$ to $\hat{\Sigma}\cup \hat{\Sigma}_1$, as discussed below. Attaching the additional $S$ operators $W_S$ to the bosonic sequential circuit $V^b$ yields the fermionic sequential circuit $V^f$. 

\subsubsection{Definition of local unitary for sequential circuits}
\begin{figure}[ht]
    \centering
    \includegraphics[width=1.0\linewidth]{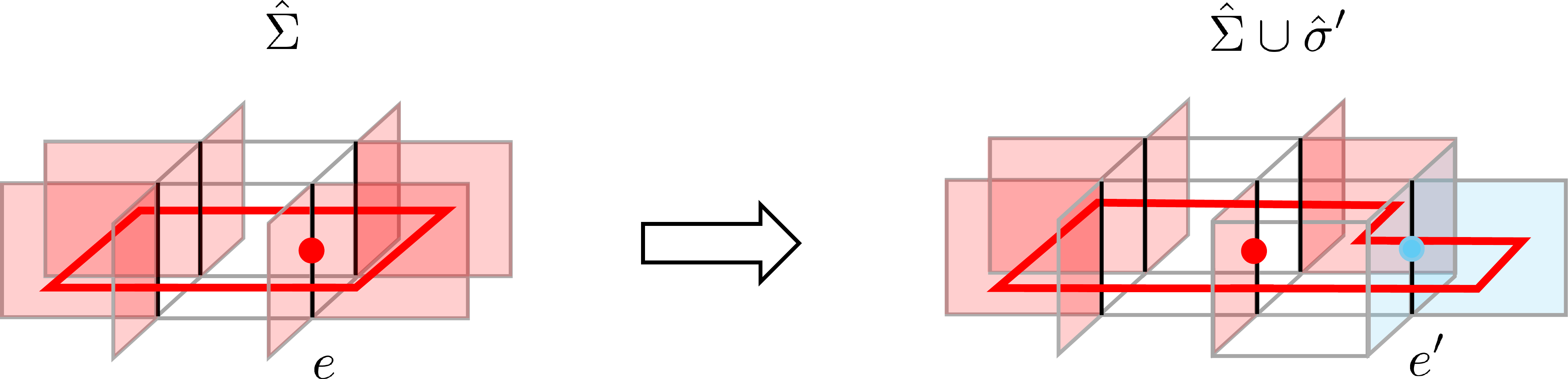}
    \caption{Left: The red bold line shows the loop excitation that violates the red faces. We focus on an edge $e$ (red dot). Right: The loop is expanded to include the new edge $e^\prime$ (blue dot), supported by the new blue faces.}
    \label{fig:cube_move_abst}
\end{figure}

\begin{figure}[ht]
    \centering
    \includegraphics[width=1.0\linewidth]{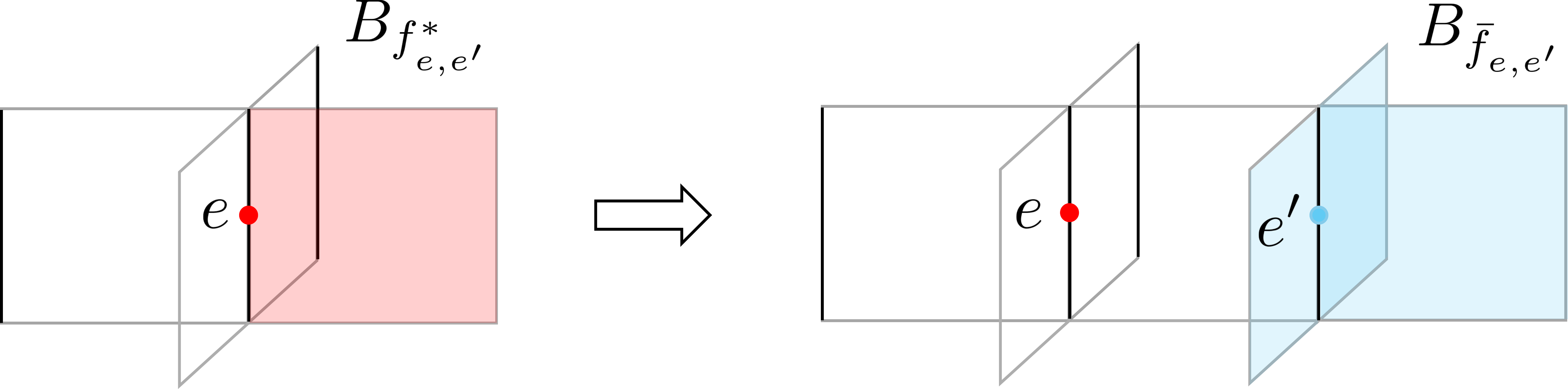}
    \caption{The red face is a support of $B_{f_{e,e^\prime}^*}$, and the blue faces are a support of $B_{\bar{f}_{e,e^\prime}}$.}
    \label{fig:cube_move_unit}
\end{figure}

We first introduce a local unitary to expand the defect by a unit open surface on the dual cube generated by a single edge (see Fig.~\ref{fig:cube_move_abst}). Let us take an edge $e$ at $\partial\hat\Sigma$, on which a magnetic flux excitation is supported (where $\partial$ denotes the boundary of a surface). Then, take an edge $e^\prime$ adjacent to $e$, and let $\hat\sigma^\prime$ be a face of the dual lattice cut by $e^\prime$. We choose $e$ such that the face $\hat\sigma^\prime$ is not contained in $\hat\Sigma$, and construct an operator that moves a defect from $\partial\hat\Sigma$ to $\partial(\hat\Sigma\cup \hat\sigma^*)$.
Let $f_{e,e^\prime}^*$ be the face containing both $e$ and $e^\prime$. We introduce the following local unitary:
\begin{align}
{U}^{b}_{e,e^\prime} &\coloneq -i \exp\left(i\frac{\pi}{2}{M}^{b}_{e,e^\prime} \right), \cr
    {M}^{b}_{e,e^\prime} &\coloneq {B}_{f_{e,e^\prime}^*}\left({X}^{A}_{e^\prime}+
     {X}^{B}_{e^\prime}\right){B}_{\bar{f}_{e,e^\prime}}\cr & \quad + {B}_{\bar{f}_{e,e^\prime}}\left({X}^{A}_{e^\prime}+{X}^{B}_{e^\prime}\right){B}_{f_{e,e^\prime}^*},
\end{align}
where the projection operator $B_{\bar{f}_{e,e^\prime}}$, supported on the faces that contain the new edge $e^\prime$ but do not contain the original edge $e$ (shown as the three blue faces in Fig.~\ref{fig:cube_move_unit}), is defined as:
\begin{align}
    B_{\bar{f}_{e,e^\prime}}\coloneq \prod_{f \ \text{s.t.}\ \partial f\ni e^\prime, \partial f\not\ni e}B_{f}.
\end{align}
Since the projections all commute with each other and are Hermitian, ${M^{b}_{e,e^\prime}}$ is Hermitian and ${U}^{b}_{e,e^\prime}$ is unitary.

\subsubsection{Action of the local unitary on the defect states}
Now we show that the above local unitary deforms the defect configurations. Let us consider the defect states $\ket{\mathcal{D}_{\hat\Sigma}^{b}}$ and  $\ket{\mathcal{D}_{\hat\Sigma \cup \hat\sigma^\prime}^{b}}$. Assume that these defect states do not include the other flux excitations. Since the flux excitation violates the flatness condition, the Hermitian operators act on these states as
\begin{align}
    {M}^{b}_{e,e^\prime}\ket{\mathcal{D}_{\hat\Sigma}^{b}} &= {B}_{f_{e,e^\prime}^*}\left({X}^{A}_{e^\prime}+{X}^{B}_{e^\prime}\right)\ket{\mathcal{D}_{\hat\Sigma}^{b}} =\ket{\mathcal{D}_{\hat\Sigma\cup \hat\sigma^\prime}^{b}},\\
    {M}^{b}_{e,e^\prime}\ket{\mathcal{D}_{\hat\Sigma\cup \hat\sigma^\prime}^{b}}&= {B}_{\bar{f}_{e,e^\prime}}\left({X}^{A}_{e^\prime}+{X}^{B}_{e^\prime}\right)\ket{\mathcal{D}_{\hat\Sigma\cup\hat\sigma^\prime}^{b}}=\ket{\mathcal{D}_{\hat\Sigma}^{b}}.
\end{align}
These relations imply
\begin{align}
    \left({M}^{b}_{e,e^\prime}\right)^2\ket{\mathcal{D}_{\hat\Sigma}^{b}} &= \ket{\mathcal{D}_{\hat\Sigma}^{b}} ,\\
    \left({M}^{b}_{e,e^\prime}\right)^2\ket{\mathcal{D}_{\hat\Sigma\cup \sigma }^{b}}&= \ket{\mathcal{D}_{\hat\Sigma\cup \sigma }^{b}}.
\end{align}
As a result, we obtain the expected actions of the local unitary on the defect states as
\begin{align}\label{eq:non-abelian_flux_unitary}
    U^{b}_{e,e^\prime}\ket{\mathcal{D}_{\hat\Sigma}^{b}} &=-i \left(\cos{\frac{\pi}{2}}+i \sin{\frac{\pi}{2}}M^{b}_{e,e^\prime}\right)\ket{\mathcal{D}_{\hat\Sigma}^{b}}= \ket{\mathcal{D}_{\hat\Sigma\cup\hat\sigma^\prime}^{b}}.
\end{align}

\subsubsection{Construction of the sequential circuit $V^{b}$}
We then define $V^{b}(\hat\Sigma,\hat\Sigma_1)$ by assigning a strict ordering to the edges $(e_1, \dots, e_N)$ as
\begin{align}
    V^{b}(\hat\Sigma,\hat\Sigma_1) = \prod_{k=1}^{N-1}{U}^{b}_{e_k,e_{k+1}},
\end{align}
where the product is path-ordered to represent a step-by-step expansion of the surface defect as follows.
Let $\hat\sigma_k$ be the unit open surface expanded by the $k$-th local unitary $U^{b}_{e_k,e_{k+1}}$. To ensure that the boundary of the intermediate surface excitation expands by steps, these unit surfaces must be sequentially connected at their boundaries such that $e_1 \in \partial \hat\Sigma \cap \partial \hat\sigma_1$ and $e_{k+1} \in \partial \hat\sigma_k \cap \partial \hat\sigma_{k+1}$. Consequently, the union of expanded surfaces is given by $\hat\Sigma_1 = \bigcup_{k=1}^{N-1} \hat\sigma_k$, which corresponds to the set of all surfaces intersected by the edges $\{e_k\}$. This completes the definition of the sequential circuit $V^b$ for bosonic magnetic fluxes, together with the fermionic ones $V^f$ in Eq.~\eqref{eq:Vf}.

\subsubsection{Unitary circuits in qudit representation}
In the $\Z_4$ qudit representation, the defect states and related sequential circuits are also defined in a similar manner.

The defect states are defined by
\begin{align}
    \ket{\widetilde{\mathcal{D}}^f_{\hat\Sigma}}&\coloneq \widetilde{S}_{\D_4}(\hat\Sigma)\ket{\text{GS}}~,\cr  \quad \ket{\widetilde{\mathcal{D}}^b_{\hat\Sigma}}&\coloneq \frac{1}{\sqrt{2}}\left(\widetilde{T}^A_X(\hat\Sigma)+\widetilde{T}^B_X(\hat\Sigma)\right)\ket{\text{GS}}~.
\end{align}

These defect configurations are deformed by sequential unitaries:
\begin{align}
    \widetilde{V}^f(\hat\Sigma,\hat\Sigma_1) = \widetilde{W}_S(\Sigma_1)\widetilde{V}^b(\hat\Sigma,\hat\Sigma_1),
    \label{eq:Vf_qudit}
\end{align}
where $\widetilde{V}^b$ is constructed from the sequential circuits $\widetilde{V}^{b}(\hat\Sigma,\hat\Sigma_1) = \prod_{k=1}^{N-1}\widetilde{U}^{b}_{e_k,e_{k+1}}$. Here we defined the local unitary $\widetilde{U}^{b}_{e,e^\prime}$ as
\begin{align}
\widetilde{U}^{b}_{e,e^\prime} &\coloneq -i \exp\left(i\frac{\pi}{2}\widetilde{M}^{b}_{e,e^\prime} \right), \cr
    \widetilde{M}^{b}_{e,e^\prime} &\coloneq \widetilde{B}_{f_{e,e^\prime}^*}\left({X}^{A}_{e^\prime}+
     {X}^{B}_{e^\prime}\right)\widetilde{B}_{\bar{f}_{e,e^\prime}}\cr & \quad + \widetilde{B}_{\bar{f}_{e,e^\prime}}\left({X}^{A\dagger}_{e^\prime}+{X}^{B\dagger}_{e^\prime}\right)\widetilde{B}_{f_{e,e^\prime}^*}.
\end{align}

Before going to the evaluation of the invariance, let us introduce the local unitary operators used later
\begin{align}
    \widetilde{\mU}^{b}_{e,e'} \coloneq  \exp\left( i\frac{\pi}{2}\cdot (-i) \widetilde{\mM}^{T_X}_{e,e'} \right),
\end{align}
where the supports of the operator are the same as $\widetilde{U}^{T_X}_{e,e'}$ as shown in Fig.~\ref{fig:cube_move_abst} and~\ref{fig:cube_move_unit}.
Here, we introduced the anti-Hermitian operator
\begin{align}
    \widetilde{\mM}^{b}_{e,e'}&\coloneq  \widetilde{B}_{f_{e,e'}^*}\left(\widetilde{X}^{A}_{e'}+\widetilde{X}^{B}_{e'}\right)\widetilde{B}_{\bar{f}_{e,e'}}\cr & \quad - \widetilde{B}_{\bar{f}_{e,e'}}\left(\widetilde{X}^{A\dagger}_{e'}+\widetilde{X}^{B\dagger}_{e'}\right)\widetilde{B}_{f_{e,e'}^*}.
\end{align}
This anti-Hermitian operator satisfies
\begin{align}
    \widetilde{\mM}^{b}_{e,e'}\ket{\widetilde{\mathcal{D}}_{\hat\Sigma}^{b}} &= \ket{\widetilde{\mathcal{D}}_{\hat\Sigma\cup\hat\sigma^\prime}^{b}},\\
    \widetilde{\mM}^{b}_{e,e'}\ket{\widetilde{\mathcal{D}}_{\hat\Sigma\cup\hat\sigma^\prime}^{b}} &= -\ket{\widetilde{\mathcal{D}}_{\hat\Sigma}^{b}},
\end{align}
which imply $\left(-i\widetilde{\mM}^{b}_{e,e'}\right)^2\ket{\widetilde{\mathcal{D}}_{\hat\Sigma}^{b}} = \ket{\widetilde{\mathcal{D}}_{\hat\Sigma}^{b}}$. Consequently, the local unitary $\widetilde{\mU}^{b}_{e,e'}$ also moves the defect of the pure magnetic loop:
\begin{align}
    \widetilde{\mU}^{b}_{e,e'}\ket{\widetilde{\mathcal{D}}_{\hat\Sigma}^{b}} &= \ket{\widetilde{\mathcal{D}}_{\hat\Sigma\cup\hat\sigma^\prime}^{b}}.
\end{align}

Furthermore, these local unitaries satisfy the following commutation relations:
\begin{align}\label{eq:comm_relations_U^tx1}
     \widetilde{U}^{b}_{e,e'} \cdot \widetilde{Z}^{A}_{e^\prime}\widetilde{Z}^{B}_{e^\prime}&=-i \widetilde{Z}^{A}_{e^\prime}\widetilde{Z}^{B}_{e^\prime}\cdot   \widetilde{\mU}^{b}_{e,e'} ,
  \end{align}
  \begin{align}\label{eq:comm_relations_U^tx2}
      \left(\widetilde{U}^{b}_{e,e'}\right)^\dagger\cdot \widetilde{Z}^{A}_{e^\prime}\widetilde{Z}^{B}_{e^\prime}&= i\widetilde{Z}^{A}_{e^\prime}\widetilde{Z}^{B}_{e^\prime}\cdot \left(\widetilde{\mU}^{b}_{e,e'}\right)^\dagger,
\end{align}
and
\begin{align}\label{eq:comm_relations_U^tx3}
    \widetilde{\mU}^{b}_{e,e'}\cdot \widetilde{Z}^{A}_{e^\prime}\widetilde{Z}^{B}_{e^\prime} &= -i\widetilde{Z}^{A}_{e^\prime}\widetilde{Z}^{B}_{e^\prime}\cdot \left(\widetilde{U}^{b}_{e,e'}\right)^\dagger,
    \end{align}
    \begin{align}\label{eq:comm_relations_U^tx4}
    \left(\widetilde{\mU}^{b}_{e,e'}\right)^\dagger \cdot \widetilde{Z}^{A}_{e^\prime}\widetilde{Z}^{B}_{e^\prime} &=i\widetilde{Z}^{A}_{e^\prime}\widetilde{Z}^{B}_{e^\prime}\cdot \widetilde{U}^{b}_{e,e'}.
\end{align}
Note that $\widetilde{U}^{b}_{e,e'}$ and $\widetilde{\mU}^{b,-}_{e,e'}$ commute with $\widetilde{Z}^{A}_{e}\widetilde{Z}^{B}_{e}$, defined on an edge $e$.

\subsubsection{24-step unitary sequence}
Let us consider the $\Z_2$ invariant for non-Abelian loop excitation. To define $\Z_2$ invariant associated to the loop excitations, we again employ the 24-step unitary sequence shown in Fig.~\ref{fig:24-sequence}, and define the Berry phase
\begin{align}
    e^{i\Theta_{f}}
    &\coloneq \bra{\mathcal{D}^f_{\hat\Sigma_0}} V^f_{014}V^f_{034}V^f_{023}{V^{f\dagger}_{014}}{{V}_{024}^{f\dagger}}V^f_{012}{{V}_{023}^{f\dagger}}\nonumber \\
\quad &\times {{V}_{013}^{f\dagger}} V^f_{024}V^f_{014}V^f_{013}{{V}_{024}^{f\dagger}}{{V}_{034}^{f\dagger}}V^f_{023}{{V}_{013}^{f\dagger}}{{V}_{012}^{f\dagger}}\nonumber \\
\quad &\times V^f_{034}V^f_{024}V^f_{012}{{V}_{034}^{f\dagger}}{{V}_{014}^{f\dagger}}V^f_{013}{{V}_{012}^{f\dagger}}{{V}_{023}^{f\dagger}}\ket{\mathcal{D}^f_{\hat\Sigma_0}},
\end{align}
where we explicitly show the supports of deformed parts in unitaries as $V^f(\hat\Sigma,\hat\Sigma_{0ij})=V^f_{0ij}\equiv W^S_{0ij}V^b_{0ij}$ and $\hat\Sigma_0$ is an initial defect configuration. This phase satisfies the desired conditions to define the invariance discussed in Sec.~\ref{sec:invariance}.

We give an outline of the explicit evaluation of the invariant:
\begin{enumerate}
    \item For simplicity of computations, we work on defect operators with $\Z_4$ qudits instead of qubits for the Hilbert space $A,B$~\cite{hsin2025higherformanomaliesimplyintrinsic}.
    \item Evaluate the relative phase between our target $e^{i\Theta_{f}}$ and the Berry phase of pure non-Abelian flux $e^{i\Theta_{b}}$.
    \item Evaluate the phase of pure non-Abelian flux $e^{i\Theta_{b}}$.
\end{enumerate}

In the first step, we replace all operators by the $\Z_4$ qudit representation for the invariance:
\begin{align}\label{eq:24-step_fermion_qudit_rep}
    e^{i\Theta_{f}}
    &= \bra{\widetilde{\mathcal{D}}^f_{\hat\Sigma_0}} \widetilde{V}^f_{014}\widetilde{V}^f_{034}\widetilde{V}^f_{023}{{\widetilde{V}}^{f\dagger}_{014}}{{\widetilde{V}}_{024}^{f\dagger}}\widetilde{V}^f_{012}{{\widetilde{V}}_{023}^{f\dagger}}\nonumber \\
\quad &\times {{\widetilde{V}}_{013}^{f\dagger}} \widetilde{V}^f_{024}\widetilde{V}^f_{014}\widetilde{V}^f_{013}{{\widetilde{V}}_{024}^{f\dagger}}{{\widetilde{V}}_{034}^{f\dagger}}\widetilde{V}^f_{023}{{\widetilde{V}}_{013}^{f\dagger}}{{\widetilde{V}}_{012}^{f\dagger}}\nonumber \\
\quad &\times \widetilde{V}^f_{034}\widetilde{V}^f_{024}\widetilde{V}^f_{012}{{\widetilde{V}}_{034}^{f\dagger}}{{\widetilde{V}}_{014}^{f\dagger}}\widetilde{V}^f_{013}{{\widetilde{V}}_{012}^{f\dagger}}{{\widetilde{V}}_{023}^{f\dagger}}\ket{\widetilde{\mathcal{D}}^f_{\hat\Sigma_0}}.
\end{align}

In the second step, we explicitly evaluate the relative phase between $e^{i\Theta_{f}}$ and $e^{i\Theta_{b}}$.
Here, the Berry phase for the pure magnetic flux is defined by
\begin{align}
    e^{i\Theta_{b}}
    &\coloneq \bra{\widetilde{\mathcal{D}}^b_{\hat\Sigma_0}} \widetilde{V}^b_{014}\widetilde{V}^b_{034}\widetilde{V}^b_{023}{{\widetilde{V}}^{b\dagger}_{014}}{{\widetilde{V}}_{024}^{b\dagger}}\widetilde{V}^b_{012}{{\widetilde{V}}_{023}^{b\dagger}}\nonumber \\
\quad &\times {{\widetilde{V}}_{013}^{b\dagger}} \widetilde{V}^b_{024}\widetilde{V}^b_{014}\widetilde{V}^b_{013}{{\widetilde{V}}_{024}^{b\dagger}}{{\widetilde{V}}_{034}^{b\dagger}}\widetilde{V}^b_{023}{{\widetilde{V}}_{013}^{b\dagger}}{{\widetilde{V}}_{012}^{b\dagger}}\nonumber \\
\quad &\times \widetilde{V}^b_{034}\widetilde{V}^b_{024}\widetilde{V}^b_{012}{{\widetilde{V}}_{034}^{b\dagger}}{{\widetilde{V}}_{014}^{b\dagger}}\widetilde{V}^b_{013}{{\widetilde{V}}_{012}^{b\dagger}}{{\widetilde{V}}_{023}^{b\dagger}}\ket{\widetilde{\mathcal{D}}^b_{\hat\Sigma_0}}.
\end{align}
Since Eq.~\eqref{eq:24-step_fermion_qudit_rep} includes an equal number of $\widetilde{W}^S_{0ij}$ and $\widetilde{W}^{S\dagger}_{0ij}$, we can evaluate the relative phase by employing commutation relations~\eqref{eq:comm_relations_U^tx1}--\eqref{eq:comm_relations_U^tx4}, and obtain 
\begin{align}\label{eq:rel_phase}
    e^{i\Theta_{f}} = (-1)\times e^{i\Theta_{b}}.
\end{align}
The detailed calculations are presented in Appendix~\ref{app:24-step_details}. 

In the final step, we evaluate the Berry phase associated with the pure magnetic flux $e^{i\Theta_{b}}$. From Eq.~\eqref{eq:non-abelian_flux_unitary}, unless there are other types of excitations, any deformation of a pure magnetic flux configuration generated by $\widetilde{V}^{b}$ does not produce a nontrivial phase. Therefore, we obtain $e^{i\Theta_{b}}=1$.

Through these three steps, we finally obtain the following result:
\begin{align}
    e^{i\Theta_{f}} = (-1)\times e^{i\Theta_{b}} = -1.
\end{align}
This result illustrates that the loop excitation has nontrivial self-statistics and it is fermionic.

\section{(3+1)D mixed topological order with a non-Abelian fermionic loop}
\label{sec:mixed}

In this section, we construct a new intrinsically mixed topological order in (3+1)D whose long-range entanglement is protected by non-Abelian fermionic loop statistics of the strong symmetry.

\subsection{Review: (3+1)D mixed topological order with a single fermionic loop}
Let us first review an intrinsically mixed topological order in (3+1)D with a single fermionic loop~\cite{hsin2025higherformanomaliesimplyintrinsic}.
We consider a cubic lattice in three spatial dimensions with a $\Z_4$ qudit residing on each edge $e$. 
A convenient stabilizer Hamiltonian realizing the (3+1)D $\Z_2$ toric code is
\begin{equation}
 \widetilde{H}_{\Z_2}
 = -\sum_e \widetilde X_e^2
 -\sum_v \left(\widetilde{A}_v + \widetilde{A}_v^\dagger\right)
 -\sum_p \widetilde{B}_p^2 .
 \label{eq:HTC_review}
\end{equation}
Here $\widetilde{A}_v$ and $\widetilde{B}_p$ are the standard $\Z_4$ toric-code vertex and plaquette terms: $\widetilde{A}_v$ is a product
of $\widetilde X_e$ operators over edges incident to a vertex $v$ (with incoming/outgoing orientation conventions),
and $\widetilde{B}_p$ is a product of $\widetilde Z_e$ operators around an elementary plaquette $p$ (with orientation-dependent
conjugations). The squared terms $\widetilde X_e^2$ and $\widetilde{B}_p^2$ energetically restrict the theory to an effective
$\Z_2$ gauge theory.

Within the stabilizer subspace of $\widetilde X_e^2$, the $\Z_2$ toric code has an emergent $\Z_2$ 1-form symmetry generated by the magnetic surface
\begin{equation}
\widetilde{S}_{b}(\hat\Sigma) \coloneq \prod_{e\subset\hat\Sigma} \widetilde X_e ,
\label{eq:Sb_review}
\end{equation}
where the product is over edges $e$ of the primal lattice intersected (cut) by $\hat\Sigma$.

\begin{figure}[htb]
\centering
\includegraphics[width=0.9\columnwidth]{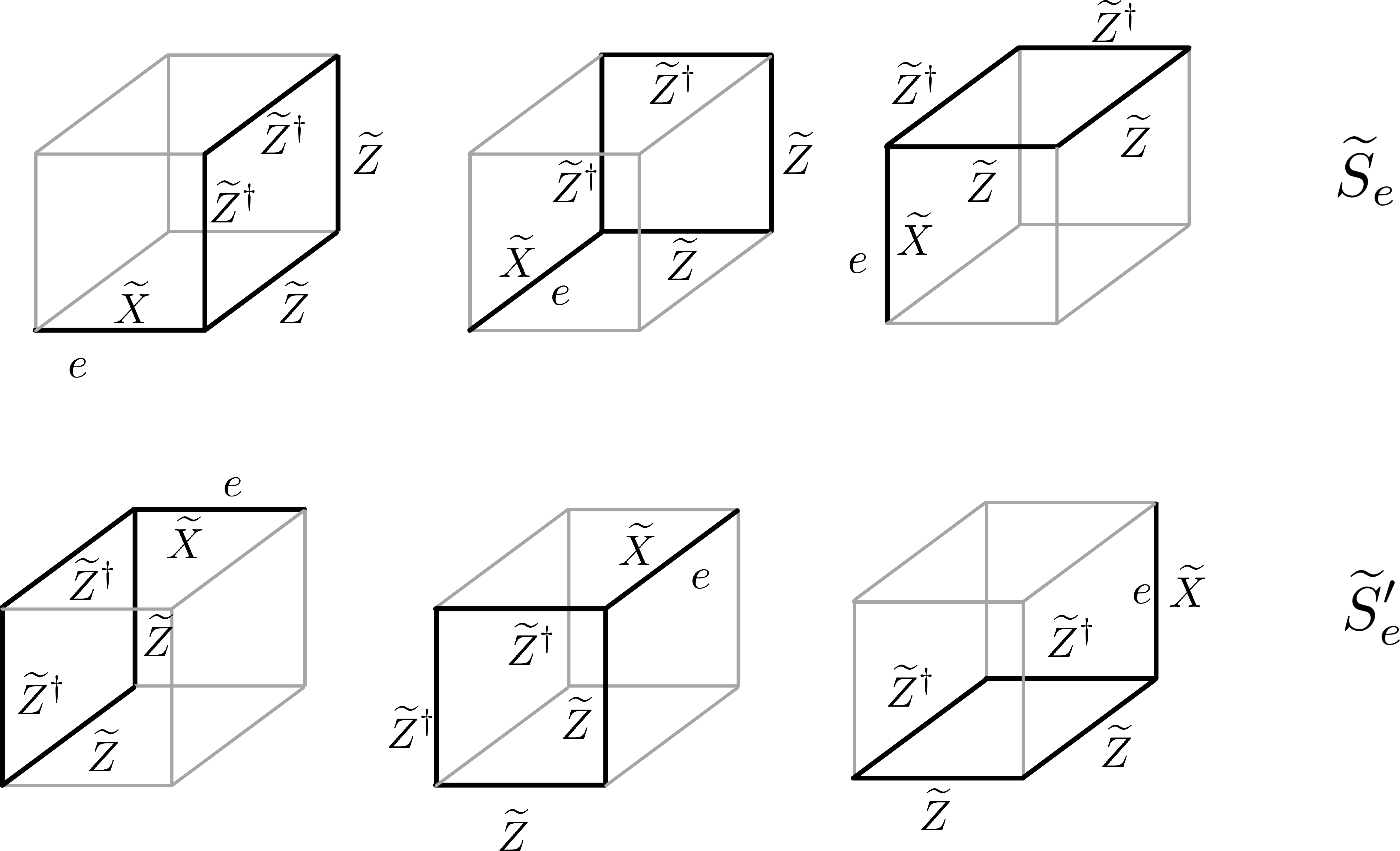}
\caption{Microscopic definitions of the local edge operators $\widetilde{S}_e$ and $\widetilde S^\prime_e$ used to construct
the fermionic membrane operator $\widetilde{S}_{\Z_2}(\hat\Sigma)$ and the local decoherence channel
$\mathcal{N}=\prod_e \mathcal{N}_e$.}
\label{fig:Se_review}
\end{figure}

Meanwhile, one may define a distinct membrane operator that generates an emergent $\Z_2$ 1-form symmetry
\begin{equation}
\widetilde{S}_{\Z_2}(\hat\Sigma) \coloneq \prod_{e\subset \hat\Sigma} \widetilde{S}_e ,
\label{eq:Sf_review}
\end{equation}
built from local edge operators $\widetilde{S}_e$ acting on the $\Z_4$ qudit on edge $e$, see Fig.~\ref{fig:Se_review}. Similar to the operators illustrated in Eq.~\eqref{eq:Se}, this $\Z_2$ 1-form symmetry again corresponds to a fermionic loop excitation.

We now define a local noise channel that incoherently applies a fermionic-loop edge operator on each edge.
Specifically, we take a product channel
\begin{equation}
\mathcal{N} = \prod_e \mathcal{N}_e,\qquad
\mathcal{N}_e(\rho) = (1-p)\,\rho + p\,\widetilde S^\prime_e\,\rho\,\widetilde S_e^{\prime\dagger} ,
\label{eq:noise_review}
\end{equation}
where $\widetilde S^\prime_e$ is a local edge operator (defined in Fig.~\ref{fig:Se_review}).

A particularly transparent regime is the maximally decohering point $p=\tfrac12$. Let the initial state be
a pure toric-code ground state $\rho_0 = \ket{\mathrm{TC}}\bra{\mathrm{TC}}$. It is often convenient to represent the mixed
state $\rho=\mathcal{N}\rho_0$ as a pure {Choi state} in a doubled Hilbert space
$\mathcal{H}_+\otimes\mathcal{H}_-$,
\begin{equation}
\ket{\rho}\rangle
= \sum_{j,k} \rho_{jk}\,\ket{j}_+\otimes \ket{k}_-^{*},
\end{equation}
where $(\cdot)^*$ denotes complex conjugation in a fixed computational basis. At $p=\tfrac12$, the Choi
state can be written compactly as
\begin{equation}
\ket{\mathcal{N}\rho_0}\rangle
=
\prod_e \left( \frac{1+\widetilde S_e^{\prime+}(\widetilde S_e^{\prime-})^{*}}{2} \right)
\ket{\mathrm{TC}}_{+}\ket{\mathrm{TC}}_{-}.
\label{eq:choi_review}
\end{equation}
This form makes the exact symmetry properties of the decohered state manifest.

A key feature of $\rho=\mathcal{N}\rho_0$ is that it retains a {strong} $\Z_2$ $1$-form symmetry
generated by $\widetilde{S}_{\Z_2}(\hat\Sigma)$. 
This {strong anomalous} $1$-form symmetry in the decohered state implies that the mixed
state $\rho=\mathcal{N}\rho_0$ is intrinsically long-range entangled as a mixed state and cannot be connected to an SRE mixed state by any finite-depth quasi-local channel.
Equivalently, it defines an intrinsically mixed-state topological order (imTO) in (3+1)D, distinct from
both (i) the original toric-code ground state (which encodes quantum logical information) and (ii) a fully
trivial mixed phase.

\subsection{(3+1)D Mixed topological order with a non-Abelian fermionic loop}

Now we explicitly construct a (3+1)D imTO with a non-Abelian fermionic loop. Let us again consider a system on a 3D cubic lattice, where we put a pair of $\Z_4$ qudits $\{\widetilde X_e^{A},\widetilde Z_e^{A}\},\{\widetilde X_e^{B},\widetilde Z_e^{B}\}$ and a single $\Z_2$ qubit $\{X_e^{C},Z_e^{C}\}$ on each edge.

We start with two copies of the imTO with a fermionic loop,
\begin{align}
   \rho_1 =  \mathcal{N}\rho_0^{A}\otimes\mathcal{N}\rho_0^{B}\otimes\ket{0}\bra{0}^{C}~,
\end{align}
where qubits $\{X_e^{C},Z_e^{C}\}$ realize the product state of $\ket{0}$. We then introduce a noise channel
\begin{equation}
\mathcal{N}' = \prod_v \mathcal{N}'_v,\qquad
\mathcal{N}'_v(\rho) = (1-p)\,\rho + p G^C_v \rho G^{C\dagger}_v~,
\label{eq:noise_swap}
\end{equation}
where the operator $G_v^{C}$ corresponds to the Gauss law gauging the $\Z_2$ symmetry swapping $A,B$ qudits
\begin{align}
  G_v^{C} = A_v^{C} \text{SWAP}_{N(v)}^{{(AB)}}\text{SWAP}_{E(v)}^{{(AB)}}\text{SWAP}_{U(v)}^{{(AB)}}
\end{align}
with $A_v^{C} $ an $X$-stabilizer of the $\Z_2$ toric code for the qubit $\{X_e^{C},Z_e^{C}\}$. The above noise channel $\mathcal{N}'$ corresponds to gauging the weak $\Z_2$ SWAP symmetry as described in Ref.~\cite{Ellison:2024svg}; gauging weak symmetry is often referred to as ``classical gauging'', since it results in a classical ensemble of gauge-field configurations for weak symmetry.

Let us consider the mixed state $\rho=\mathcal{N}'\rho_1$ with the maximal decoherence $p=1/2$. Then the decohered state $\rho$ has a strong symmetry generated by a non-Abelian fermionic loop operator \eqref{eq:SD8_qudit}. This non-Abelian fermionic loop carries the nontrivial Berry phase invariant, and protects intrinsic long-range entanglement of the mixed state. We note that aside from the non-Abelian fermionic loop excitation, the resulting mixed topological order also has a $\Z_2$ electric particle excitation for the $\Z_2^C$ gauge field. This generates a strong $\Z_2$ 2-form symmetry.

\section{Discussions}
\label{sec:discussions}
We close by highlighting several future directions. Although sequential-circuit invariants provide a transparent diagnostic for an important subclass of anomalies, they do not yet complete the definition of non-invertible 't Hooft anomalies. While it is expected that the Berry phase invariants would capture a wide class of 't Hooft anomalies including those corresponding to non-Abelian braiding invariants \cite{kaidi2023symmetrytftsanomaliesnoninvertible, inamura202321dsymmetrytopologicalorderlocalsymmetric, kawagoe2020microscopic} or Frobenius-Schur indicators \cite{Zhang:2023wlu,Cordova:2023bja, Hsin2026coset}, it is unlikely that e.g., the anomaly of $\mathbb{Z}_2$ Kramers-Wannier duality is characterized by the Berry phase invariant.
It would be valuable to formulate a complete lattice anomaly theory for non-invertible symmetries, and to clarify which anomalies are detectable purely from sequential circuits. 

Our construction of non-Abelian fermionic loops should generalize to higher-dimensional anyonic excitations. For example, consider a (4+1)D theory with symmetry $\mathbb{Z}_3\ltimes \mathbb{Z}^{(1)}_9$, where the $\mathbb{Z}^{(1)}_9$ 1-form symmetry carries anomaly $\frac{2\pi}{9}B_2B_2B_2$ with $B_2$ the 2-form $\Z_9$ background, which corresponds to nontrivial anyonic membrane statistics~\cite{Feng2026anyonic}. $\mathbb{Z}_3$ then acts by the automorphism $n\mapsto 4n \ (\mathrm{mod}\ 9)$, which leaves the anomaly invariant. Gauging the $\mathbb{Z}_3$ symmetry then yields a non-Abelian anyonic membrane from the resulting coset 1-form symmetry. Exploring such extended operators and their non-Abelian statistics is an interesting direction for future work.

\begin{acknowledgments}
RK thanks Yu-An Chen, Po-Shen Hsin, Kansei Inamura, Abhinav Prem, Sahand Seifnashri, and Matthew Yu for discussions.
RK thanks Yu-An Chen, Yitao Feng, Po-Shen Hsin, Abhinav Prem, Hanyu Xue and Matthew Yu for related collaborations. RK thanks Yu-An Chen and Po-Shen Hsin for comments on a draft. RK is supported by the Department of Applied Physics, the University of Tokyo.
Y.H.\ was supported by JST, CREST Grant Number JPMJCR24I3, and by JSPS KAKENHI Grants No.~JP25K01002, No.~JP25K24464, and No.~JP24K00630.
\end{acknowledgments}

\bibliography{bibliography.bib}

\onecolumngrid

\vspace{0.3cm}

\newpage

\appendix

\section{Invariance of Berry phases under local deformations}\label{sec:berry_phases}

In this appendix, we establish the invariance of Berry phase invariants under possible local deformations. As described in the main text, the Berry phase is generally expressed as a sum of phase factors $\theta(\mathcal{D}, \mathcal{D}')$,
\begin{align}
    \Theta = \sum_{a,\mathcal{D},\mathcal{D}'} \eps(\Sigma,a;\mathcal{D}, \mathcal{D}') \theta(\Sigma,a;\mathcal{D}, \mathcal{D}')~,
\end{align}
with $\eps(\Sigma,a;\mathcal{D}, \mathcal{D}')\in\Z$. Then, the necessary and sufficient condition for $\Theta$ to define an invariant is expressed as a set of linear constraints satisfied by the integer coefficients $\epsilon$. Here we focus on the condition arising from the local deformations of the movement operators presented in Eq.~\eqref{eq:epsilon locality constraint} in the main text.

We require invariance under possible local deformations of a movement operator $V(\Sigma, \mathcal{D}_1,\mathcal{D}_2)$ near its boundary. As described in the main text, we only consider the local deformations such that the locality of the deformation is preserved by the conjugation action of movement operators. 
    
    In general, we consider the deformation in the form of
    \begin{align}
        V'(\Sigma,a;\mathcal{D}_1,\mathcal{D}_2) = O_p(\Sigma,a;\mathcal{D}_1|_{D_{r,p}},\mathcal{D}_2|_{D_{r,p}}) V(\Sigma,a;\mathcal{D}_1,\mathcal{D}_2)~,
        \end{align}
        with some local operator $O_p$ supported at $\eps$-neighborhood of the point $p$ within the boundary $\partial\Sigma$, such that the locality of $O_p$ is preserved by generic movement operators $V(\Sigma,a;\mathcal{D},\mathcal{D}'), V(\mathcal{D})$:
        \begin{align}
\mathcal{X}_{a;\mathcal{D},\mathcal{D}'}(O_p):= V(\Sigma,a;\mathcal{D},\mathcal{D}')^\dagger O_p V(\Sigma,a;\mathcal{D},\mathcal{D}')~, \quad \mathcal{X}_{\mathcal{D}}(O_p):= V(\mathcal{D})^\dagger O_p V(\mathcal{D})~,
\end{align}
        is again a local operator at $p$.
        
        Here, $O_p$ can depend on the configurations of defects through those in the $r$-neighborhood of $p$. All movement operators $V(\Sigma,a;\mathcal{D},\mathcal{D}')$ with generic configurations $\mathcal{D},\mathcal{D}'$ supported at $\Sigma$ is modified by $O_p(\Sigma,a;\mathcal{D}|_{D_{r,p}},\mathcal{D}'|_{D_{r,p}})$; one can see that such modification is consistent with the constraint \eqref{eq:locality of V}.
        
        Accordingly, the states possibly get modified by a local unitary, 
        \begin{align}
            \ket{\mathcal{D}}' = U_{p}(\mathcal{D}|_{D_{r,p}})\ket{\mathcal{D}}~,
        \end{align}
        which can again depend on $\mathcal{D}$ through its configuration in the $r$-neighborhood of $p$. The locality of $ U_{p}(\mathcal{D}|_{D_{r,p}})$ is again preserved by generic movement operators $V(\Sigma,a;\mathcal{D},\mathcal{D}'), V(\mathcal{D})$.
        
Let us study how the Berry phase $\theta(\Sigma,a;\mathcal{D}_1,\mathcal{D}_2)$ is shifted by these redefinitions.
The Berry phase $\theta'$ after the redefinition is given via
\begin{align}
    U_{2,p}^\dagger O_p V(\Sigma,a;\mathcal{D}_1,\mathcal{D}_2)U_{1,p}\ket{\mathcal{D}_1} = e^{i\theta'(\Sigma,a;\mathcal{D}_1,\mathcal{D}_2)}\ket{\mathcal{D}_2}~,
\end{align}
where we wrote $U_p(\mathcal{D}_i|_p)$ as $U_{i,p}$.
By acting $V(\Sigma,a^*;\mathcal{D}_2, \mathcal{D}_1)$ on both sides of the above equation, we get
\begin{align}
    ^{[\Sigma,a;\mathcal{D}_1,\mathcal{D}_2]}(U_{2,p}^\dagger O_p ) U_{1,p} \ket{\mathcal{D}_1} = e^{i\theta'(\Sigma,a;\mathcal{D}_1,\mathcal{D}_2)-i\theta(\Sigma,a;\mathcal{D}_1,\mathcal{D}_2)}\ket{\mathcal{D}_1}~,
\end{align}
where $^{[\Sigma,a;\mathcal{D}_1,\mathcal{D}_2]}O:= V(\Sigma,a;\mathcal{D}_1,\mathcal{D}_2)^\dagger OV(\Sigma,a;\mathcal{D}_1,\mathcal{D}_2)$ denotes conjugation action of $V$.
Recalling that $\ket{\mathcal{D}} = V(\mathcal{D})\ket{0}$ we get
        \begin{align}
\bra{0} ^{[\mathcal{D}_1]}(^{[\Sigma,a;\mathcal{D}_1,\mathcal{D}_2]}(U_{2,p}^\dagger O_p ) U_{1,p}) \ket{0} = e^{i\theta'(\Sigma,a;\mathcal{D}_1,\mathcal{D}_2)-i\theta(\Sigma,a;\mathcal{D}_1,\mathcal{D}_2)}~,
\end{align}
where $^{[\mathcal{D}]}O:= V(\mathcal{D})^\dagger OV(\mathcal{D})$ is the conjugation action.
Now, the lhs only depends on $\mathcal{D}_1, \mathcal{D}_2$ through their configurations in the $r$-neighborhood of $p$. Therefore the phase shift has the form of
\begin{align}
    \theta'(\Sigma,a;\mathcal{D}_1, \mathcal{D}_2) = \theta(\Sigma,a;\mathcal{D}_1, \mathcal{D}_2) + \phi(\Sigma,a;\mathcal{D}_1|_{D_{r,p}}, \mathcal{D}_2|_{D_{r,p}})~.
\end{align}
The invariance under such shifts corresponds to the linear constraints
\begin{align}
    \sum_{\substack{\mathcal{D}_1,\mathcal{D}_2, \\ (\mathcal{D}_1,\mathcal{D}_2)|_{D_{r,p}} = (\mathcal{D}_{1}|, \mathcal{D}_{2}|)}}(\epsilon(\Sigma_{a;\mathcal{D}_1,\mathcal{D}_2},a;\mathcal{D}_1, \mathcal{D}_2)-\epsilon(\Sigma_{a;\mathcal{D}_1,\mathcal{D}_2},a^*;\mathcal{D}_2, \mathcal{D}_1)) = 0~,
    \label{eq:epsilon locality constraint app}
\end{align}
where the sum means that we sum over defects with fixed configurations $(\mathcal{D}_{1}|, \mathcal{D}_{2}|)$ at the $r$-neighborhood of $p$. This constraint corresponds to Eq.~\eqref{eq:epsilon locality constraint} in the main text.

\section{Detailed calculations for Eq.~\eqref{eq:rel_phase}}\label{app:24-step_details}
Let us first consider commutation relations with $\widetilde{W}_S$ and $\widetilde{V}_{b}$.
Suppose $\widetilde{W}^{S}_{{0jk}}$ and $\widetilde{V}^{b}_{{0lm}}$ have an intersection on their boundaries, i.e., $\partial \Sigma_{0jk} \cap\partial \hat{\Sigma}_{0lm}\neq \emptyset$. Then, the commutation relation between these operators is given as
\begin{align}
    \widetilde{V}^{b}_{0lm}\cdot \widetilde{W}^{S}_{0jk} 
    &=\omega\left(\widetilde{V}^{b}_{0lm}\cdot W^{S}_{0jk} \right)\cdot  \widetilde{W}^{S}_{0jk}\widetilde{V}^{b}_{{0lm}}(0jk)  ,\\
     \widetilde{V}^{b}_{{0lm}}(0jk)&\coloneq \mathcal{P}\left[\widetilde{V}^{b}_{\hat{\Sigma}_{0lm}\setminus (\partial \Sigma_{0jk} \cap\partial \hat{\Sigma}_{0lm})} \times \prod_{e^\prime\in (\partial \Sigma_{0jk}\cap\hat{\Sigma}_{0lm})} \widetilde{\mU}^{b}_{e(e^\prime),e^\prime}\right],
\end{align}
where $e = e(e^\prime)$ is the edge in $\hat{\Sigma}_{0lm}$ that constitutes a face where $e^\prime$ is defined, and $\mathcal{P}[\cdot]$ denotes the proper path-ordered product to represent a step-by-step expansion of the surface defect. Here, we denote the exchange phase factor as $\omega\left(\widetilde{V}^{b}_{0lm}, \widetilde{W}^{S}_{0jk} \right)\in \{\pm1,\pm i\}$. 
When the Hermitian conjugate ${\widetilde{W}^{S\dagger}_{{0jk}}}$ is considered, we denote the modified operator as $\widetilde{V}^{b}_{{0lm}}(0jk^{\dagger})$. Furthermore, for multiple surfaces $\Sigma_{0jk}, \Sigma_{0ab},\ldots$ associated with $\widetilde{W}^S$, we simply denote the sequentially modified operator as $\widetilde{V}^{b}_{0lm}(0jk,0ab,\ldots)$, provided that these surfaces are mutually disjoint on the boundary.

Similarly, we also have
\begin{align}
    \widetilde{V}^{b\dagger}_{0lm}\cdot \widetilde{W}^{S}_{0jk} 
    &=\omega\left(\widetilde{V}^{b\dagger}_{0lm}\cdot \widetilde{W}^{S}_{0jk} \right)\times \widetilde{W}^{S}_{0jk}\cdot \mathcal{P}\left[\widetilde{V}^{b\dagger}_{\hat{\Sigma}_{0lm}\setminus (\partial \Sigma_{0jk} \cap\partial \hat{\Sigma}_{0lm})} \times \prod_{e^\prime\in (\partial \Sigma_{0jk}\cap\hat{\Sigma}_{0lm})} \left(\widetilde{\mU}^{b}_{e(e^\prime),e^\prime}\right)^\dagger\right].
\end{align}

Note that from Eqs.~\eqref{eq:comm_relations_U^tx1},~\eqref{eq:comm_relations_U^tx2},~\eqref{eq:comm_relations_U^tx3} and~\eqref{eq:comm_relations_U^tx4}, phase factors satisfy the following relations:
\begin{align}\label{eq:phase_relation}
   \omega\left(\widetilde{V}^{b}_{{0lm}}, \widetilde{W}^{S}_{{0jk}}\right)\cdot \omega\left(\widetilde{V}^{b\dagger}_{{0lm}}, \widetilde{W}^{S}_{{0jk}}\right) = 1,\\
      \omega\left(\widetilde{V}^{b}_{{0lm}}, \widetilde{W}^{S\dagger}_{{0jk}}\right)\cdot \omega\left(\widetilde{V}^{b\dagger}_{{0lm}}, \widetilde{W}^{S\dagger}_{{0jk}}\right) = 1.
\end{align}

Before proceeding to the explicit evaluation, we also note several commutation relations that follow from the supports of the operators. The operators $\{\widetilde{V}^f_{012}, \widetilde{V}^f_{014}, \widetilde{V}^f_{023}, \widetilde{V}^f_{034}\}$ all commute with each other. Furthermore, $\widetilde{V}^f_{024}$ commutes with both $\widetilde{V}^f_{023}$ and $\widetilde{V}^f_{034}$. Similarly, $\widetilde{V}^f_{013}$ commutes with both $\widetilde{V}^f_{012}$ and $\widetilde{V}^f_{023}$.

Using these relations, we now obtain
\begin{align}
    e^{i\Theta_{f}} =  &c_1\times \bra{\widetilde{\mathcal{D}}^f_{\hat\Sigma_0}}  \widetilde{V}^{b}_{014}\widetilde{V}^{b}_{034}\widetilde{V}^{b}_{023}{V^{b\dagger}_{014}}{\widetilde{V}_{024}^{b\dagger}}\widetilde{V}^{b}_{012}{\widetilde{V}_{023}^{b\dagger}}{\widetilde{V}_{013}^{b\dagger}}\nonumber \\
\quad &\times \widetilde{V}^{b}_{024}(012^{\dagger},013)\cdot \widetilde{V}^{b}_{014}(013)\cdot \widetilde{V}^{b}_{013}\cdot {\widetilde{V}_{024}^{b\dagger}}(014^{\dagger},012^{\dagger})\cdot{\widetilde{V}_{034}^{b\dagger}}\widetilde{V}^{b}_{023}{\widetilde{V}_{013}^{b\dagger}}{\widetilde{V}_{012}^{b\dagger}}\nonumber\\
\quad &\times \widetilde{V}^{b}_{034}(013)\cdot \widetilde{V}^{b}_{024}(013,014^{\dagger})\cdot\widetilde{V}^{b}_{012}\cdot {\widetilde{V}_{034}^{b\dagger}}(013)\cdot {\widetilde{V}_{014}^{b\dagger}}(013)\cdot \widetilde{V}^{b}_{013}{\widetilde{V}_{012}^{b\dagger}}{V_{023}^{b\dagger}}\ket{\widetilde{\mathcal{D}}^f_{\hat\Sigma_0}},
\end{align}
where we denoted the phase $c_1$ as
\begin{align}
c_1 \coloneq  &\omega\left(\widetilde{V}^{b}_{024}, \widetilde{W}^{S\dagger}_{{012}}\right)\cdot \omega\left(\widetilde{V}^{b}_{{024}}, \widetilde{W}^{S}_{{013}}\right)\cdot \omega\left(\widetilde{V}^{b}_{{014}}, \widetilde{W}^{S}_{{013}}\right)\cdot \omega\left(\widetilde{V}^{b\dagger}_{{024}}, \widetilde{W}^{S\dagger}_{{014}}\right)\cdot \omega\left(\widetilde{V}^{b\dagger}_{{024}}, \widetilde{W}^{S\dagger}_{{012}}\right) \nonumber\\
    &\times\omega\left(\widetilde{V}^{b}_{{034}}, \widetilde{W}^{S}_{{013}}\right)\cdot \omega\left(\widetilde{V}^{b}_{{024}}, \widetilde{W}^{S}_{{013}}\right)\cdot \omega\left(\widetilde{V}^{b}_{{024}}, \widetilde{W}^{S\dagger}_{{014}}\right) \cdot \omega\left(\widetilde{V}^{b\dagger}_{{034}}, \widetilde{W}^{S}_{{013}}\right)\cdot \omega\left(\widetilde{V}^{b\dagger}_{{014}}, \widetilde{W}^{S}_{{013}}\right),
\end{align}
and used $(\hat{\Sigma}_{024}\cap\Sigma_{013})\cap(\hat{\Sigma}_{024}\cap\Sigma_{012})=\emptyset$ and  $(\hat{\Sigma}_{024}\cap\Sigma_{013})\cap(\hat{\Sigma}_{024}\cap\Sigma_{014})=\emptyset$ for applying the relation~(\ref{eq:phase_relation}).
Since $\widetilde{V}^{b}_{{024}}, \widetilde{W}^S_{{013}}$ has the same support of qudits, the overall phase can be evaluated as
\begin{align}
c_1=\omega\left(\widetilde{V}^{b}_{{024}}, \widetilde{W}^{S}_{{013}}\right)\cdot \omega\left(\widetilde{V}^{b}_{{024}}, \widetilde{W}^{S}_{{013}}\right)= (-i)^2 = -1.
\end{align}

\begin{figure}[t]
    \centering
    \includegraphics[width=0.8\linewidth]{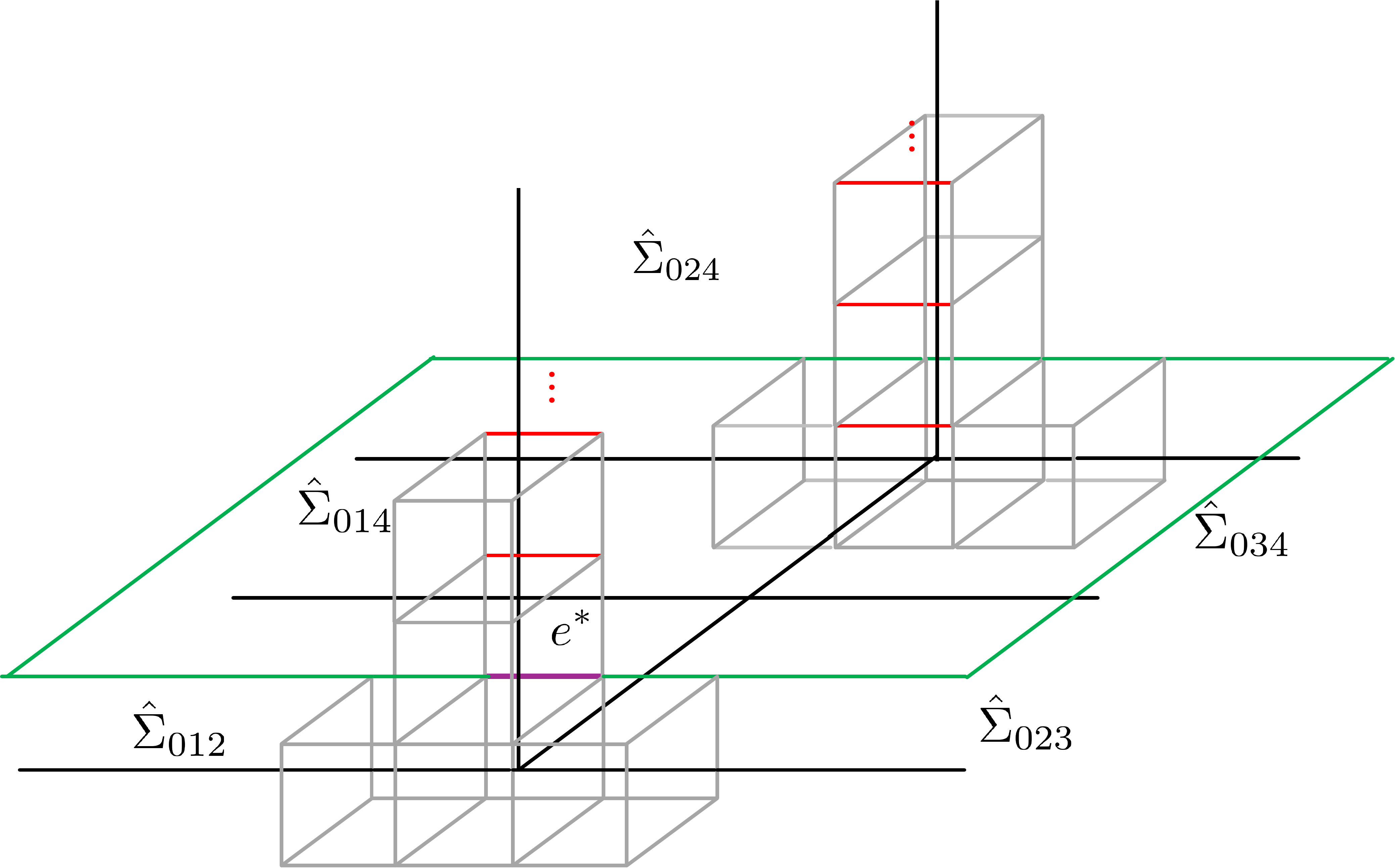}
    \caption{Black lines show edges of dual surfaces representing boundaries of $\hat\Sigma_{0ij}$, and grey cubes denote the primal cubic lattice. The green line denotes the edges where $\widetilde{W}^S(\Sigma_0)$ is defined, and red edges show the part of the supports of $\widetilde{V}^{b}_{024}$. These two surface operators overlap only on the purple edge $e^*$.}
    \label{fig:supports_overlap}
\end{figure}

Next, we also rewrite the expectation value of the 24-sequence by $\ket{\widetilde{\mathcal{D}}^{b}_{\hat\Sigma_0}}$ instead of $\ket{\widetilde{\mathcal{D}}^f_{\hat\Sigma_0}}$:
\begin{align}\label{eq:relative_phase}
&\bra{\widetilde{\mathcal{D}}^f_{\hat\Sigma_0}}  \widetilde{V}^{b}_{014}\widetilde{V}^{b}_{034}\widetilde{V}^{b}_{023}{\widetilde{V}^{b\dagger}_{014}}{\widetilde{V}_{024}^{b\dagger}}\widetilde{V}^{b}_{012}{\widetilde{V}_{023}^{b\dagger}}{\widetilde{V}_{013}^{b\dagger}}\nonumber \\
\quad &\times \widetilde{V}^{b}_{024}(012^{\dagger},013)\cdot \widetilde{V}^{b}_{014}(013)\cdot \widetilde{V}^{b}_{013}\cdot {\widetilde{V}_{024}^{b\dagger}}(014^{\dagger},012^{\dagger})\cdot{\widetilde{V}_{034}^{b\dagger}}\widetilde{V}^{b}_{023}{\widetilde{V}_{013}^{b\dagger}}{\widetilde{V}_{012}^{b\dagger}}\nonumber\\
\quad &\times \widetilde{V}^{b}_{034}(013)\cdot \widetilde{V}^{b}_{024}(013,014^{\dagger})\cdot \widetilde{V}^{b}_{012}\cdot {\widetilde{V}_{034}^{b\dagger}}(013)\cdot {\widetilde{V}_{014}^{b\dagger}}(013)\cdot \widetilde{V}^{b}_{013}{\widetilde{V}_{012}^{b\dagger}}{\widetilde{V}_{023}^{b\dagger}}\ket{\widetilde{\mathcal{D}}^f_{\hat\Sigma_0}}\nonumber\\
&=c_2\times\bra{\widetilde{\mathcal{D}}^{b}_{\hat\Sigma_0}}  \widetilde{V}^{b}_{014}\widetilde{V}^{b}_{034}\widetilde{V}^{b}_{023}{\widetilde{V}^{b\dagger}_{014}}{\widetilde{V}_{024}^{b\dagger}}(\sigma^*)\widetilde{V}^{b}_{012}{V_{023}^{b\dagger}}{\widetilde{V}_{013}^{b\dagger}}\nonumber \\
\quad &\times \widetilde{V}^{b}_{024}((012^{\dagger},013),\sigma^*)\cdot \widetilde{V}^{b}_{014}(013)\cdot \widetilde{V}^{b}_{013}\cdot {\widetilde{V}_{024}^{b\dagger}}((014^{\dagger},012^{\dagger}),\sigma^*)\cdot{\widetilde{V}_{034}^{b\dagger}}\widetilde{V}^{b}_{023}{V_{013}^{b\dagger}}{\widetilde{V}_{012}^{b\dagger}}\nonumber\\
\quad &\times \widetilde{V}^{b}_{034}(013)\cdot \widetilde{V}^{b}_{024}(013,014^{\dagger},\sigma^*)\cdot \widetilde{V}^{b}_{012}\cdot {\widetilde{V}_{034}^{b\dagger}}(013)\cdot {\widetilde{V}_{014}^{b\dagger}}(013)\cdot \widetilde{V}^{b}_{013}{\widetilde{V}_{012}^{b\dagger}}{\widetilde{V}_{023}^{b\dagger}}\ket{\widetilde{\mathcal{D}}^{b}_{\hat\Sigma_0}},
\end{align}
with the relative phase $c_2$:
\begin{align}
    c_2\coloneq \omega\left(\widetilde{V}^{b\dagger}_{024}, \widetilde{W}^{S}_{\sigma^*}\right)\cdot \omega\left(\widetilde{V}^{b}_{024}(012^\dagger,013), \widetilde{W}^{S}_{\sigma^*}\right)\cdot \omega\left(\widetilde{V}^{b\dagger}_{024}(014^\dagger,012^\dagger), \widetilde{W}^{S}_{\sigma^*}\right)\cdot \omega\left(\widetilde{V}^{b}_{024}(013,014^\dagger), \widetilde{W}^{S}_{\sigma^*}\right).
\end{align}
Here, we used $W^S(\Sigma_0)$ in the initial state commute with all $\widetilde{V}^{b}_{0ij}$ except $\widetilde{V}^{b}_{024}$ and their Hermitian conjugation on an edge $e^*$ cut by dual surface $\sigma^*$. The detailed overlap of the support of $\widetilde{V}^{b}_{024}$ and $\widetilde{W}^S(\Sigma_0)$ is shown in Fig.~\ref{fig:supports_overlap}. 
We also defined $\widetilde{V}^{b}_{024}((012^{\dagger},013),\sigma^*)$ and ${\widetilde{V}_{024}^{b\dagger}}((014^{\dagger},012^{\dagger}),\sigma^*)$ as
\begin{align}
    \widetilde{V}^{b}_{024}(012^{\dagger},013)\cdot \widetilde{W}^S(\Sigma_0) &= \omega\left(\widetilde{V}^{b}_{024}(012^\dagger,013), \widetilde{W}^{S}_{\sigma^*}\right)\cdot \widetilde{W}^S(\Sigma_0)\widetilde{V}^{b}_{024}((012^{\dagger},013),\sigma^*),\\
    \widetilde{V}^{b}_{024}((012^{\dagger},013),\sigma^*)&\coloneq \mathcal{P}\left[\widetilde{V}^{b}_{\hat\Sigma_{024}\setminus\{(\partial\Sigma_{012}\cap\partial\hat\Sigma_{024}\setminus\sigma^*)\cup (\partial\Sigma_{013}\cap\partial\hat\Sigma_{024})\}}\cdot \widetilde{U}^{b}_{e(e^*),e^*}\right],
\end{align}
and
\begin{align}
    {\widetilde{V}_{024}^{b\dagger}}(014^{\dagger},012^{\dagger})\cdot \widetilde{W}^S(\Sigma_0) &= \omega\left(V^{b}_{024}(014^\dagger,012^\dagger), \widetilde{W}^{S}_{\sigma^*}\right)\cdot \widetilde{W}^S(\Sigma_0){\widetilde{V}_{024}^{b\dagger}}((014^{\dagger},012^{\dagger}),\sigma^*),\\
    {\widetilde{V}_{024}^{b\dagger}}((014^{\dagger},012^{\dagger}),\sigma^*)&\coloneq \mathcal{P}\left[\widetilde{V}^{b\dagger}_{\hat\Sigma_{024}\setminus\{(\partial\Sigma_{012}\cap\partial\hat\Sigma_{024}\setminus\sigma^*)\cup (\partial\Sigma_{014}\cap\partial\hat\Sigma_{024})\}}\cdot \left(\widetilde{U}^{b}_{e(e^*),e^*}\right)^\dagger\right].
\end{align}

Now, relative phases are evaluated as
\begin{align}
    c_2= i \cdot i \cdot (-i)\cdot (-i)= 1.
\end{align}
Since the eight modified unitaries such as ${\widetilde{V}_{024}^{b\dagger}}(\sigma^*)$ in Eq.~\eqref{eq:relative_phase} do not change the relative phase from the original ones, we obtain $e^{i\Theta_{f}}= c_1c_2\times e^{i\Theta_{b}} =  (-1)\times e^{i\Theta_{b}}$.

\end{document}